\documentclass[aps,prx,reprint,superscriptaddress,nofootinbib,longbibliography]{revtex4-1}

\usepackage{amsmath,amssymb, amsthm,amstext}
\usepackage{natbib}
\usepackage{graphicx}
\usepackage{color}
\usepackage{array, enumerate}
\usepackage{bm}
\usepackage{multirow}
\usepackage[breaklinks,colorlinks,citecolor=blue]{hyperref}
\usepackage{braket}
\usepackage{txfonts}
\usepackage{physics}
\usepackage{placeins}
\usepackage[normalem]{ulem}

\def\be{\begin{equation}}
\def\ee{\end{equation}}

\begin{document}

\title{Captured are circularized:  A relativistic \\ treatment  of extreme mass ratio inspirals crossing accretion disks}
\author{Yuhe Zeng}
\email{zengyuhe@sjtu.edu.cn}
\affiliation{Tsung-Dao Lee Institute, Shanghai Jiao-Tong University, Shanghai, 520 Shengrong Road, 201210, People’s Republic of China}
\affiliation{School of Physics \& Astronomy, Shanghai Jiao-Tong University, Shanghai, 800 Dongchuan Road, 200240, People’s Republic of China}

\author{Zhen Pan}
\email{zhpan@sjtu.edu.cn}
\affiliation{Tsung-Dao Lee Institute, Shanghai Jiao-Tong University, Shanghai, 520 Shengrong Road, 201210, People’s Republic of China}
\affiliation{School of Physics \& Astronomy, Shanghai Jiao-Tong University, Shanghai, 800 Dongchuan Road, 200240, People’s Republic of China}

\begin{abstract}
 A small body orbiting around an accreting massive object and periodically crossing its accretion disk is a common configuration in astrophysics. 
 In this work, we study the secular evolution of extreme mass-ratio inspirals (EMRIs), where a stellar-mass object (SMO)—such as a star or a stellar-mass black hole (sBH)—collides with the accretion disk surrounding a central supermassive black hole (SMBH). Using a perturbation method applied to EMRI geodesics, we find  the following: (1) the disk tends to align the SMO regardless of its initial inclination $\iota$ relative to the disk, (2) the final orbital eccentricity of an SMO captured by the disk is low, although the eccentricity can temporarily increase when the initial inclination $\iota$ is large and the SMO is an sBH, (3) through collisions with the accretion disk alone, only a small fraction of sBHs that initially lie close to both the SMBH and the disk can be captured within the typical disk lifetime of active galactic nuclei. Two-body scatterings among SMOs in the nuclear stellar cluster play an essential role, randomly kicking sBHs toward the disk and significantly boosting the capture rate.
\end{abstract}
\date{\today}

\maketitle

\section{Introduction}

Astrophysical systems in which a small body orbits around an accreting massive central object appear in a wide range of environments, e.g.,  (i) planetesimals interacting with circumstellar material around  white dwarfs or young planets embedded in protoplanetary disks, where gas drag or disk-planet torques drive orbital migration and damp eccentricity and inclination~\cite{OConnor2020,annurev:/content/journals/10.1146/annurev-astro-081811-125523,refId1}; (ii) compact stellar-mass objects (SMOs) or stars interacting with the dense gas of active galactic nuclei (AGN) disks, which can trap, align, and migrate small objects and facilitate hierarchical mergers~\cite{Yang_2019,Peng_2025,10.1111/j.1745-3933.2011.01132.x,10.1111/j.1365-2966.2006.11155.x,10.1093/mnras/stw2260,Tagawa_2020}; (iii) SMOs around supermassive black holes (SMBHs) interacting with transient accretion disks formed in tidal disruption events and producing x-ray  quasi-periodic eruptions~\cite{Xian_2021,Linial_2023,refId2,10.1093/mnras/stad2616,PhysRevD.109.103031,PhysRevD.110.083019,Zhou_2025,Zhou:2025auz}.
Together, these systems provide a unified physical setting in which an orbiting secondary interacts with the central massive object itself or with its ambient environment, leading to secular modifications of the orbit and distinctive radiative signatures. A substantial body of work has focused on how such interactions drive the secular evolution of orbiting secondaries. Depending on the environment, long-term orbital changes may arise from gravitational radiation, dissipative gas torques, dynamical friction~\cite{refId0,spieksma2025gripdiskdraggingcompanion}, or impulsive perturbations during disk crossings~\cite{OConnor2020,Wang2024,spieksma2025gripdiskdraggingcompanion,Rowan_2025,10.1093/mnras/staa3004,10.1093/mnras/stad1295,Secunda_2021,MacLeod:2019jxd}. Over timescales much longer than the orbital period, the small effects produced by these forces accumulate into a net secular drift of the orbital elements, corresponding to the usual adiabatic evolution of the system.

 One class of such astrophysical systems, extreme mass-ratio inspiral (EMRI), typically consisting of a central SMBH and an orbiting SMO, which can be a stellar companion or a compact remnant such as a stellar-mass black hole (sBH), satisfying a mass ratio $q\equiv m_{\rm SMO}/M_{\rm SMBH}\lesssim 10^{-4}$, is among the primary gravitational-wave sources targeted by the future space-based missions LISA~\cite{baker2019laserinterferometerspaceantenna}, TianQin~\cite{10.1093/ptep/ptaa114}, and Taiji~\cite{10.1093/ptep/ptaa083}. Their long-lived, relativistic dynamics encode detailed information about the strong-field spacetime and thereby enable precision tests of general relativity (GR) and measurements of SMBH properties~\cite{AmaroSeoane2018,Barack:2018yvs,Gair:2012nm}. Although EMRIs are often treated as vacuum systems, realistic AGNs contain dense stellar clusters and accretion disks, within which compact objects may reside or migrate~\cite{PhysRevD.75.024034,10.1111/j.1365-2966.2006.11155.x,PhysRevD.103.103018,PhysRevD.104.063007,Pan_2021,PhysRevD.105.083005,10.1093/mnras/stad749}.
 In such environments, SMOs in EMRIs may experience non-negligible matter-induced forces, including aerodynamic drag, hydrodynamical torques, and gas dynamical friction, all of which may significantly modify their orbital evolution relative to vacuum inspirals. Such systems, often termed ``wet EMRIs"~\cite{PhysRevD.104.063007}, can deviate in both secular orbital-parameter evolution and inspiral timescales, affecting not only the astrophysical interpretation of their dynamics but also the accuracy of waveform templates required for LISA data analysis~\cite{PhysRevD.84.024032,10.1093/mnras/staa3976,10.1093/mnras/stz1026}.

Formation of wet EMRIs in AGN disks by capturing SMOs on misaligned orbits has been investigated in detail \cite{PhysRevD.105.083005}. 
Two-body scatterings randomly kick SMOs onto low-inclination orbits, and  density waves excited by low-inclination SMOs along with dynamical friction efficiently capture the SMOs onto the AGN disk. Density waves drive the SMOs to migrate inward as well as damping their orbital eccentricities~\cite{Tanaka_2002,Tanaka_2004}. As a result, wet EMRIs in the LISA sensitivity band are expected to be of low eccentricity. On the other hand, there are also recent studies~\cite{10.1093/mnras/stae1239,PhysRevD.111.084006} of eccentric wet EMRIs and their scientific potentials motivated by previous studies showing that SMO-disk collisions may increase the orbital eccentricity in some parameter space~\cite{Secunda_2021,Wang2024}.

\begin{figure*}[t]
    \centering
    \includegraphics[scale=0.45]{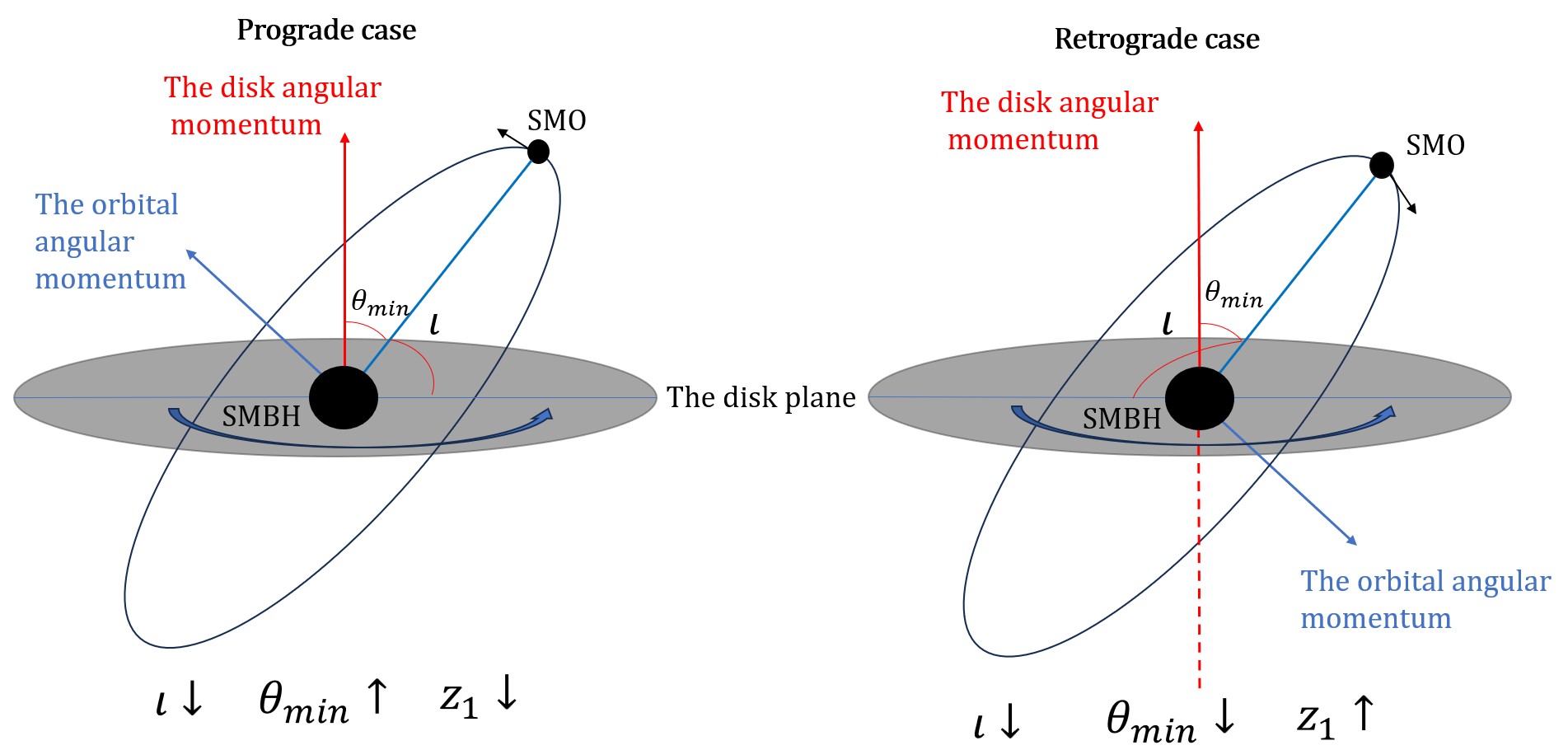}
    \caption{Schematic illustration of the wet EMRIs with SMO-disk collision scenario. The relations between the labeled parameters are $z_{1}\equiv\cos\theta_{\rm min}=\sin\iota$, where $\Vec{L}_{\rm disk}$ is the disk angular momentum, and $\Vec{L}_{\rm orbit}$ is the orbital angular momentum of the SMO, the angle between these two vectors is equivalent to the orbital inclination, $\cos^{-1}(\hat{L}_{\rm disk}\vdot\hat{L}_{\rm orbit})=\iota$. $0<\iota<90^{\circ}$ corresponds to the prograde cases, while $90^{\circ}<\iota<180^{\circ}$ corresponds to the retrograde cases. In both cases $\iota$ will decrease, which will be verified in this work.}
    \label{fig:EMRI scheme}
\end{figure*}

In this work, we investigate the orbital evolution induced by SMO-disk collisions using a perturbation method applied to EMRI geodesics. Specifically, we model the motion of the SMO as a forced geodesic in Schwarzschild spacetime and consider its repeated crossings of a geometrically thin accretion disk located on the equatorial plane (see Fig.~\ref{fig:EMRI scheme}). To quantify the secular evolution of the orbit, we derive the equations governing the variations of the principal orbital elements $a$, $e$, and $\iota$, directly with respect to the proper time $\tau$. By applying an adiabatic approximation procedure over the fast orbital phases, we extract the net secular evolution driven by SMO-disk collisions and determine the associated characteristic timescales for orbital shrinkage, inclination damping, and also the tendency of eccentricity evolution. 
Once the SMO is captured by the disk, its orbit no longer undergoes discrete disk crossings but instead enters an embedded phase. In that regime, the orbital evolution is expected to be governed mainly by continuous exchange of energy and angular momentum with the disk through gas torques, rather than by the impulsive collisions considered in this work. Several previous works have developed relativistic frameworks for EMRIs in this embedded phase~\cite{9src-p7sp,duque2025extrememassratioinspiralsrelativisticaccretion,dyson2026spiraldensitywavestorque,g83s-jdld}, showing that strong-field effects can substantially modify the torque behavior relative to Newtonian expectations. In the present work, however, we focus only on the precapture disk-crossing stage and do not model the subsequent embedded evolution.

This paper is organized as follows. 
In Sec.~\ref{sec:basic framework}, we introduce the evolution equations for the orbital elements, the adiabatic orbit-averaging scheme used to obtain their secular behavior, and also summarize the two force models considered in this work.
In Sec.~\ref{sec:results}, we present the secular evolution of the orbital parameters under the two force models and derive the scaling relations that govern the characteristic timescales, and compare our findings with previous studies, extending the timescale estimates using the adopted disk model and scaling relations.
We conclude this work with  Sec.~\ref{sec:conclusion}. In Appendixes~\ref{appendix:form of acceleration}, \ref{appendix:coefficients}, and \ref{appendix:average Integral2}, we provide details of the key formulas used in this work. Throughout this work, we use geometrized units with $G=c=1$.

\section{Basic framework}
\label{sec:basic framework}
\subsection{Forced EMRI motion around a SMBH}\label{sec:forced motion}
Within the GR framework, the motion of a freely falling test particle is simply described by a geodesic of the background spacetime. 
However,  the geodesic motion is perturbed by external forces in nonvacuum environments and the equation of motion (EOM) is written as 
\begin{equation}
    \dv[2]{x^{\mu}}{\tau}+\Gamma^{\mu}_{\alpha\beta}\dv{x^{\alpha}}{\tau}\dv{x^{\beta}}{\tau}=f^{\mu},
    \label{eq: forced geodesic}
\end{equation}
where $x^{\mu}$ is the four-coordinate of the SMO around the central SMBH, and $f^{\mu},\ \tau$ denote the four-acceleration and proper time, respectively. In the vacuum environment, the external force vanishes,  $f^{\mu}=0$.  
The form of $f^{\mu}$ in the reference of the zero-angular-momentum observers (ZAMOs) has been given in Ref.~\cite{PhysRevD.83.044037}, $f^{\mu}$ was assumed to depend linearly on the spatially projected component of the relative four-velocity $u_{\perp}^{\mu}$, and in order to enforce the orthogonality condition, additional correction terms associated with $u_{\rm ZAMO}^{\mu}$ were also included. In this work, following a similar consideration based on the relative four-velocity components, we replace $u^{\mu}_{\rm ZAMO}$ with the four-velocity of gas in the accretion disk, $u^{\mu}_{\rm disk}$. With the constraints of $u^{\mu}u_{\mu}=-1$ and equivalently, $f^{\mu}u_{\mu}=0$, the form of $f^{\mu}$ is
\begin{equation}
    f^{\mu}=-\gamma\left(u^{\mu}+\frac{u^{\mu}_{\rm disk}}{u_{{\rm disk}\ \nu} u^{\nu}}\right),
    \label{eq:force form}
\end{equation}
where $\gamma$ denotes the interaction (damping) coefficient that encodes the strength of the force exerted on the SMO.


In the case of an SMO interacting with an accretion disk, the disk is assumed to follow nearly Keplerian rotation around the central SMBH and to be geometrically thin. The force exerted on the SMO therefore depends not only on the local disk properties but also on the relative velocity between the SMO and the disk material. In this work, we incorporate both the spatial dependence of the interaction coefficient and the actual relative velocity determined by the Keplerian motion of the disk material within the thin-disk approximation. Starting from the density profile of the disk, we adopt an axisymmetric configuration in which the thin-disk density distribution is expressed as a function of the radial coordinate $r$ and the polar angle $\theta$, assuming a Gaussian density profile in the vertical direction:
\begin{equation}
    \rho_{\rm gas}(r,\ \theta)= \rho_{\rm g} \exp\left[-\frac{(\theta-\pi/2)^2}{2(H_{\rm disk}/r)^2}\right],
    \label{eq: rho gas disk}
\end{equation}
where $\rho_{\rm g}$ is the reference density of the disk and $H_{\rm disk}$ denotes the disk scale height to the midplane. This distribution effectively confines the disk material to the vicinity of the equatorial plane when $H_{\rm disk}\ll r$. Consequently, in our calculations the SMO interacts significantly with the gas only when it passes near the midplane. Moreover, since the present analysis is restricted to a finite radial interval and aims to isolate the dynamical impact of disk crossings rather than to model the global disk structure, we omit any additional radial dependence beyond the exponential angular profile introduced above.
Under the thin-disk approximation, the disk matter is confined to $\theta=\pi/2$, and the corresponding four-velocity components are
\begin{equation}
\begin{aligned}
     &u_{\rm disk}^{r}=u_{\rm disk}^{\theta}=0, \\ 
    &g_{\mu\nu}u_{\rm disk}^{\mu}u^{\nu}_{\rm disk}=-1, \\
    &u_{\rm disk}^{\phi}=\Omega_{K}u_{\rm disk}^{t},
\end{aligned}
\end{equation}
where 
\begin{equation}
    \Omega_{K}^{\pm}=\pm\frac{\sqrt{M_{\bullet}}}{r^{3/2}\pm a_{\bullet}\sqrt{M_{\bullet}}}
\end{equation}
denotes the Keplerian angular velocity in Kerr spacetime with $a_{\bullet}$ the spin parameter of the central SMBH.
The ``+" branch corresponds to prograde rotation, i.e., the angle between the disk angular momentum and the SMBH spin is less than $90^{\circ}$, whereas the ``-" branch corresponds to retrograde rotation. In the present work, which is carried out in Schwarzschild spacetime, the sign is determined by the vertical projection of the orbital angular momentum of the SMO onto the disk angular momentum. By substituting $u^{\mu}_{\rm ZAMO}$ with $u^{\mu}_{\rm disk}$, the four-acceleration for the SMO-disk collision scenario is derived, with the $r,\ \theta,\ \phi$ components presented in Appendix~\ref{appendix:form of acceleration}.

\subsection{Damping coefficient $\gamma$}\label{ref:disk coefficient}
In this Subsection, we compute the damping coefficient $\gamma$ for both star-disk collisions and sBH-disk collisions. The coefficient $\gamma$ depends on several factors, including the interaction type, the size of the SMO, the gas density and the scale height of the disk.\par
According to the thin-disk density profile in Eq.~\eqref{eq: rho gas disk}, the effects of SMO-disk collisions are strongly localized near the equatorial plane, $\theta=\pi/2$. In this region, where $\abs{\theta-\pi/2}\ll 1$, one may approximate $r^2(\theta-\pi/2)^2/H_{\rm disk}^2\approx r^2\cos^2\theta/H_{\rm disk}^2$.
In terms of disk surface density $\Sigma_{\rm g}$ and disk height to the midplane $H_{\rm disk}$, the reference gas density $\rho_{\rm g}$ in Eq.~\eqref{eq: rho gas disk} is estimated by treating the thin disk as having an effective vertical thickness $2H_{\rm disk}$ and a uniform density within this narrow region as  $\rho_{\rm g}\approx \Sigma_{\rm g}/2 H_{\rm disk}$. The damping coefficient is then formulated as 
\begin{equation}
    \gamma(r,\ \theta)=\gamma_{\rm g}\exp\left[-\frac{(\theta-\pi/2)^2}{2(H_{\rm disk}/r)^2}\right],
\end{equation}
where $\gamma_{\rm g}$ is the reference damping coefficient apart from the spatial-distribution dependence.\par
For star-disk collisions, the perturbation force is dominated by aero-drag force~\cite{Wang2024}
\begin{equation}
    \boldsymbol{F}_{\rm aero} = -\pi R_{\star}^2\rho_{\rm g}v_{\rm rel}\boldsymbol{v}_{\rm rel},
    \label{eq:aero drag}
\end{equation}
where $\boldsymbol{v}_{\rm rel}\equiv \boldsymbol{v}-\boldsymbol{v}_{\rm g}$ denotes the relative velocity between the star and the local gas in the disk,  $v_{\rm rel}\equiv\abs{\boldsymbol{v}_{\rm rel}}$, $R_{\star}$ is the stellar radius and $\rho_{\rm g}$ is the gas density of the disk. Consequently, the reference damping coefficient $\gamma_{\rm g}$ can be written in the numerical form as 
\begin{equation}
\begin{aligned}
    \gamma_{\rm g} &=  \pi R_{\star}^2\rho_{\rm g} v_{\rm rel}m_{\star}^{-1} \\ 
    &= 2.55\times 10^{-8}\left(\frac{R_{\star}} {R_{\odot}}\right)^{2}\left(\frac{m_{\star}}{M_{\odot}}\right)^{-1}\\
    &\cdot\left( \frac{v_{\rm rel}}{0.1 c}\right)
    \left(\frac{\rho_g}{ 
    10^{5}\ \mathrm{g\cdot cm^{-2}}/3M_\bullet}\right) \left[M_{\bullet}^{-1}\right]\ .
    \label{eq:aero-drag force gamma}
\end{aligned}
\end{equation}

For sBH-disk collisions, the effective radius of an sBH with mass $m_{\bullet}$ is given by its horizon radius $R_{\rm H}=2m_{\bullet}$, which is much smaller than that of a stellar object with the same mass, and the perturbation force is dominated by dynamical friction~\cite{Wang2024,Ostriker_1999}
\begin{equation}
    \bm{F}_{\rm dyn}=-\mathcal{I}\frac{4\pi(Gm_{\bullet})^2\rho_{\rm g}}{v_{\rm rel}^{3}}\bm{v}_{\rm rel},
    \label{eq:Fdyn}
\end{equation}
where the coefficient $\mathcal{I}$ is related with Mach number $\mathcal{M}\equiv v_{\rm rel}/c_{s}$.
In this work, we note that $\mathcal{I}\sim \mathcal{O}(1)$ in the relevant parameter regime and mainly rescales the interaction strength. We therefore set $\mathcal{I}=1$, for simplicity, which does not affect the qualitative orbital-evolution behavior discussed below\footnote{ For typical orbital scales $r\sim 10^2\ M_{\bullet}$ in this work, the motion of the SMO is expected to be supersonic, in which case $\mathcal{I}\sim\Lambda\sim\mathcal{O}(1)$, where $\Lambda$ denotes the Coulomb logarithm~\cite{Wang2024}.}. The reference damping coefficient $\gamma_{\rm g}$ associated with the acceleration dominated by dynamical friction can be written in the numerical form as
\begin{equation}
\begin{aligned}
    \gamma_{\rm g}&=  \mathcal{I} \frac{4\pi G^2 m_{\bullet}\rho_{\rm g}}{v_{\rm rel}^{3}} \approx \frac{4\pi G^2 m_{\bullet}\rho_{\rm g}}{v_{\rm rel}^{3}}\\
    &= 4.62\times 10^{-15}\ \left(\frac{m_{\bullet}}{M_{\odot}}\right)\left(\frac{v_{\rm rel}}{0.1\ c}\right)^{-3} \left(\frac{\rho_g}{ 
    10^{5}\ \mathrm{g\cdot cm^{-2}}/3M_\bullet}\right) \left[M_{\bullet}^{-1}\right].
    \label{eq:dynamical friction gamma}
\end{aligned}
\end{equation}

As discussed in~\cite{Wang2024}, the appropriate interaction model--either aero-drag force~\eqref{eq:aero drag} or dynamical friction~\eqref{eq:Fdyn}--is dictated by the ratio of the SMO velocity relative to the disk gas, $v_{\rm rel}$, to its surface escape velocity, $v_{\rm esc}$. For $v_{\rm rel}/v_{\rm esc}>1$, ram-pressure drag dominates and the aero-drag prescription applies,  which is the relevant regime for the stellar objects considered in this work. For sBHs, on the other hand, $v_{\rm esc} \sim c$ implies $v_{\rm rel}/v_{\rm esc}<1$, rendering dynamical friction the appropriate description. We also note that, for stellar objects, the separation between the two regimes may become ambiguous when the orbit approaches coplanarity with the disk, as $v_{\rm rel}\rightarrow 0$. In this limit, the condition $v_{\rm rel}>v_{\rm esc}$ may eventually cease to hold, and the interaction may be gradually dominated by dynamical friction. However, this transition occurs only near the end of the capture process and has a negligible impact on the capture timescale estimated in this work. We therefore adopt the aero-drag force model throughout the orbital evolution of the stellar cases presented below.

\par
For clarity, the reference damping coefficient $\gamma_{\rm g}$ is expressed as the combination of a constant baseline value $\gamma_{0}$ and a velocity-dependent contribution arising from the relative motion. Specifically, for aero-drag force, $\gamma_{\rm g}=\gamma_{0}\ (v_{\rm rel}/0.1\ c)$, whereas for dynamical friction, $\gamma_{\rm g}=\gamma_{0}\ (v_{\rm rel}/0.1\ c)^{-3}$. The relative velocity is computed from 
\begin{equation}
    v_{\rm rel}=\sqrt{1-\frac{1}{(u_{ \mu}u^{\mu}_{\rm disk})^2}} \ .
\end{equation}
For comparison, Ref.~\cite{PhysRevD.105.084041} also derived an explicit relativistic expression of dynamical friction, which can be expressed in the form
\begin{equation}
    f^{\mu}=4\pi\rho G^2 m_\bullet \Lambda\frac{[2(u\cdot U)^2-1]^2}{[(u\cdot U)^2-1]^{\frac{3}{2}}}(u\cdot U)\left[u^{\mu} + \frac{U^{\mu}}{u\cdot U}\right] \ ,
    \label{eq:dynamical friction fully relativistic form}
\end{equation}
where $\rho$ is the proper mass density of the medium, and $U^{\mu}$ is the four-velocity of the medium. In terms of the general form in Eq.~\eqref{eq:force form}, the prefactor $\gamma$ is formulated as
\begin{equation}
\begin{aligned}
    \gamma &= 4\pi\rho G^2m_\bullet\Lambda\frac{[2(u\cdot U)^2-1]^2}{[(u\cdot U)^2-1]^{\frac{3}{2}}}\abs{u\cdot U}\\
    &= 4\pi\rho G^2m_\bullet\Lambda\frac{(1+v_{\rm rel}^2)^2}{1-v_{\rm rel}^2}\frac{1}{v_{\rm rel}^{3}},
    \label{eq:gamma dynamical friction}
\end{aligned}
\end{equation}
which reduces to the form in Eq.~\eqref{eq:dynamical friction gamma}, after identifying $U^{\mu}$ with $u^{\mu}_{\rm disk}$, $\rho$ with the local gas density
and $\mathcal{I} =\Lambda\frac{(1+v_{\rm rel}^2)^2}{1-v_{\rm rel}^2} $.  For the cases investigated in this work, $v_{\rm rel}$ is not extremely relativistic, and $\mathcal{I}$ remains an $\mathcal{O}(1)$ factor.


\subsection{Orbital evolution: Osculating orbit}\label{sec:parameter evolution}
There are effectively seven degrees of freedom in the orbital dynamics of an SMO in curved spacetime: four coordinate components and four velocity components subject to one normalization constraint $u^{\alpha} u_{\alpha}=-1$. To parametrize this motion, we introduce a set of five orbital elements, corresponding to the degrees of freedom beyond the $t$ and $\phi$ coordinates, denoted by the conserved parameter set $\mathbf{I}=\{p,\ e,\ z_{1}\}$, and phases $\psi_{r},\ \psi_{\theta}$. These parameters are related to the radial and polar motions in Boyer-Lindquist coordinates through
\begin{equation}
\begin{aligned}
    &r=\frac{p}{1+e\cos\psi_{r}}\ , \\
    &\cos\theta =z_{1}\cos\psi_{\theta}\ .
    \label{eq:r theta Kepler}
\end{aligned}
\end{equation}
Here, $p$ is the semilatus rectum, $e$ is the eccentricity, and $z_{1}\equiv \cos\theta_{\rm min}$, where $\theta_{\rm min}$ is the minimum polar angle reached by the orbit.
Analytic expression of $z_1$ in terms of conserved quantities of geodesics in Kerr spacetime can be found in \cite{Schmidt:2002qk}. 
In the Schwarzschild limit considered in this work, one may equivalently define the orbital inclination with respect to the disk plane by $\cos\theta_{\rm min}=\sin\iota$. The ranges $0<\iota<90^{\circ}$ and $90^{\circ}<\iota<180^{\circ}$ then correspond to prograde and retrograde orbits with respect to the disk angular momentum, respectively.

For simplicity, we consider a Schwarzschild SMBH of mass $M_{\bullet}$ in this work. The spacetime metric is
\begin{equation}
    ds^2=-\left(1-\frac{2M_{\bullet}}{r}\right)dt^2+\left(1-\frac{2M_{\bullet}}{r}\right)^{-1}dr^2+r^2\left(d\theta^2+\sin^2\theta\ d\phi^2\right) \ .
    \label{eq:Schwarzschild metric}
\end{equation}
Owing to the spherical symmetry of the Schwarzschild background, for spherically symmetric perturbations, the polar parameter $z_{1}$ remains constant, and the orbital plane may be chosen as the equatorial plane without loss of generality, leaving only a single radial phase parameter $\psi_{r}$ to characterize the motion. However, in the SMO-disk collision scenario, the breaking of spherical symmetry by the existence of an equatorial disk requires an additional polar phase parameter $\psi_{\theta}$ to fully characterize the orbital evolution.\par

Following Ref.~\cite{Fujita_2009}, the equations of motion of a test particle around a Schwarzschild black hole are written as 
\begin{equation}
\begin{aligned}
    \left(\dv{r}{\tau}\right)^{2}&=E^{2}-\left(1-\frac{2M_{\bullet}}{r}\right)\left(1+\frac{J^{2}}{r^{2}}\cdot\frac{1}{1-z_{1}^{2}}\right)\ , \\
    \left(\dv{\theta}{\tau}\right)^{2}&=\frac{J^{2}}{r^{4}\sin^{2}\theta}\cdot\frac{1}{1-z_{1}^{2}}\cdot(z_{1}^{2}-\cos^{2}\theta)\ , \\ 
     \dv{t}{\tau}&=E\left(1-\frac{2M_{\bullet}}{r}\right)^{-1}\ ,\\
      \dv{\phi}{\tau}&=\frac{J}{r^{2}}\cdot\frac{1}{1-\cos^{2}\theta}\ ,
\end{aligned}
\label{eq:four equations of motion}
\end{equation}
where $E,\ J$ are the conservation quantities, energy and $z$ component of the orbital angular momentum respectively. In the following derivations, we set $M_{\bullet}=1$ for simplicity.

Based on Eqs.~\eqref{eq:four equations of motion}, 
the frequencies of the unperturbed geodesic motion for phases $\psi_{r},\ \psi_{\theta}$ can be obtained as
\begin{equation}
\begin{aligned}
     \dv{\psi_{r}}{\tau}&:=\omega_{r}=\frac{1}{(\partial r/\partial\psi_{r})}\dv{r}{\tau}=(1+e\cos\psi_{r})^{2}\sqrt{\frac{p-6-2e\cos\psi_{r}}{p^{3}(p-3-e^{2})}}\ , \\
    \dv{\psi_{\theta}}{\tau}&:=\omega_{\theta}=\frac{1}{(\partial\theta/\partial\psi_{\theta})}\dv{\theta}{\tau}=\frac{(1+e\cos\psi_{r})^{2}}{p\sqrt{p-3-e^{2}}}\ .
    \label{eq:dpsidtau and dchidtau}
\end{aligned}
\end{equation}
Notice that $\omega_{\theta}$ depends solely on $\psi_{r}$. By integrating $\dv{\psi_{\theta}}{\psi_{r}}=\omega_{\theta}/\omega_{r}$, the relation between two phases is found as
\begin{equation}
    \psi_{\theta}(\psi_{r})=2\sqrt{\frac{p}{p-6-2e}}F\left(\frac{\psi_{r}}{2}\ \Bigg|\ \frac{4e}{6+2e-p}\right)+ \psi_{\theta\ \rm ini},
\end{equation}
where the function $F$ is the incomplete elliptic integral of the first kind,
\begin{equation}
    F(\phi\ |\ m)=\int_{0}^{\phi}\frac{\rm d\theta}{\sqrt{1-m\sin^2\theta}}.
\end{equation}

The forced motion of an SMO can be obtained by directly  evolving the EOM [Eq.~\eqref{eq: forced geodesic}]. Then the orbital parameters $p,\ e,\ z_{1}$ can be extracted by inverting Eqs.~\eqref{eq:four equations of motion} \cite{Fujita_2009}. However, as the evolution timescale increases far exceeding the orbital period, the computational cost and the difficulty of precision control for such method rise rapidly, limiting the applicability to long-time evolution problems.

This motivates the development of faster evolution methods that average over the rapid oscillations on the orbital timescale and focus on the secular evolution of the orbital parameters, which can be divided into two main steps. First, the EOM is reformulated in terms of orbital elements $\{p,\ e,\ z_{1}, \ \psi_r, \ \psi_\theta\}$ using the method of the osculating orbit. Second, applying the adiabatic approximation to the reformulated EOM and extracting secular evolution of orbital elements as in the following Sec.~\ref{sec:secular evolution}.

Following Refs.~\cite{PhysRevD.83.044037,PhysRevD.77.044013}, the osculation condition can be written as  
\begin{equation}
\begin{aligned}
     x^{\mu}(\tau)&=x_{\rm G}^{\mu}[\tau,\ I_{0}(\tau)]\ , \\ 
    \dv{x^{\mu}}{\tau}&=\pdv{x_{\rm G}^{\mu}}{\tau}[\tau,\ I_{0}(\tau)] \ ,
\end{aligned}
\label{eq:x xG}
\end{equation}
where $x^{\mu}_{\rm G}$ is the geodesic with orbital parameter set $I_{0}=\{p,\ e,\ z_{1},\ \psi_{0},\ \chi_{0}\}$. Here $\psi_{0}$ and $\chi_{0}$ are the “initial” values of  $\psi_r$ and $\psi_\theta$, respectively. With definition $\psi_{r}=\psi-\psi_{0}$ one can decompose the phase evolution in the radial direction as a geodesic part $\psi$ governed by Eqs.~\eqref{eq:dpsidtau and dchidtau}, and a perturbation induced shift in the “initial” phase $\psi_{0}$. In the same way, the phase evolution in the polar direction is decomposed as two parts, $\psi_{\theta}=\chi-\chi_{0}$. Defined in this way, parameters
$I_{0}$ remain conserved for unperturbed geodesic motion, however, in the presence of perturbations, it undergoes a slow secular drift superposed with oscillatory variations on the orbital timescale.  


With the osculation condition, one can reformulate the EOM ~\eqref{eq: forced geodesic} in terms of parameters $I_0$.
At each instant $\tau$, the worldline is treated as a geodesic tangent to the true trajectory, with instantaneous orbital element set $I_{0}(\tau)$.
Considering four-velocity $u^\mu$ along with the first osculation condition in  Eq.~\eqref{eq:x xG}, one can find \cite{PhysRevD.77.044013}
\begin{equation}
    u^{\mu}:=\dv{x^{\mu}}{\tau}=\pdv{x^{\mu}_{\rm G}}{\tau}+\pdv{x^{\mu}_{\rm G}}{I_{0}^a}\dv{I_{0}^a}{\tau}=\dv{x^{\mu}_{\rm G}}{\tau},
    \label{eq:osculation 0}
\end{equation}
which in combination with the second osculation condition in  Eq.~\eqref{eq:x xG} immediately leads to 
\begin{equation}
    \pdv{x^{\mu}_{\rm G}}{I_{0}^{a}}\dv{I_{0}^a}{\tau}=0 \ .
    \label{eq:osculating a}
\end{equation}

Considering four-acceleration $\mathrm{d}u^\mu/\mathrm{d}\tau$ along with  Eqs.~(\ref{eq:osculation 0}-\ref{eq:osculating a}) and 
the geodesic definition of $x^{\mu}_{\rm G}$ 
\begin{equation}
    \pdv[2]{x_{\rm G}^{\mu}}{\tau}+\Gamma^{\mu}_{\alpha\beta}\pdv{x_{\rm G}^{\alpha}}{\tau}\pdv{x_{\rm G}^{\beta}}{\tau}=0 \ ,
\end{equation}
the EOM~\eqref{eq: forced geodesic} can be rewritten as 
\begin{equation}
     \dv[2]{x^{\mu}}{\tau}=-\Gamma^\mu_{\alpha\beta} u^\alpha u^\beta + f^\mu = \pdv[2]{x_{\rm G}^{\mu}}{\tau}+f^{\mu}\ .
     \label{eq:dv2xmutaufmu}
\end{equation}

On the other hand, the same quantity can also be expressed as
\begin{equation}
    \dv[2]{x^{\mu}}{\tau}=\pdv[2]{x_{\rm G}^{\mu}}{\tau}+\pdv{I_{0}^a}\left(\dv{x^{\mu}_{\rm G}}{\tau}\right)\dv{I_{0}^a}{\tau}\ ,
    \label{eq:dv2xmutau}
\end{equation}
using Eqs.~\eqref{eq:x xG} and ~\eqref{eq:osculating a}. 
With Eqs.~\eqref{eq:dv2xmutaufmu} and~\eqref{eq:dv2xmutau},  the forced EOM~\eqref{eq: forced geodesic} can be written as
\begin{equation}
    \pdv{I_{0}^a}\left(\dv{x^{\mu}_{\rm G}}{\tau}\right)\dv{I_{0}^a}{\tau}=f^{\mu}.
    \label{eq:osculating b}
\end{equation}

The final step is to isolate individual components $d I_0^a/d\tau$ from Eqs.~\eqref{eq:osculating a} and~\eqref{eq:osculating b}, which contains 
eight equations. By imposing the orthogonality condition $f^{\mu}u_{\mu}=0$, and evolving $t$ and $\phi$ explicitly using the geodesic expressions evaluated along the instantaneous orbit~\cite{PhysRevD.83.044037}, three equation components, i.e., $\mu=t,\ \phi$ components of Eq.~\eqref{eq:osculating a} and the $\mu=t$ component of Eq.~\eqref{eq:osculating b}, can be removed. 
As a result, the evolution of the five orbital parameters with respect to $\tau$ can be written generally as
\begin{equation}
    \dv{I_{0}}{\tau}=C^{(I_{0})}_rf^{r}+C^{(I_{0})}_{\theta}f^{\theta}+C^{(I_{0})}_{\phi}f^{\phi},
    \label{eq:parameter evolution ode}
\end{equation}
with the detailed expressions of coefficients $C^{(I_{0})}_{r,\theta,\phi}$ shown in Appendix~\ref{appendix:coefficients}.

\subsection{Secular evolution}\label{sec:secular evolution}
In systems subject to weak perturbations, the orbital motion is well approximated by a geodesic on short timescales, while the perturbations accumulate over many orbital periods and induce a slow secular evolution of the orbital parameters. Such separation between the short orbital period and the long inspiral timescale justifies the use of the adiabatic approximation. For star-disk collisions considered in this work, the SMO interacts with the thin disk twice per orbital period, and each encounter produces a small perturbation, allowing this method to reliably determine the secular evolution.\par

Consider a perturbed system characterized by $N$ phases $\boldsymbol{\psi}=\{\psi_{1},\ \psi_{2},\cdots\psi_{N}\}$ and a set of orbital parameters $\mathbf{I}=\{I_1,\ I_2,\cdots I_{M}\}$ that would remain conserved in the absence of perturbations. For a function $f=f(\boldsymbol{\psi},\ \mathbf{I})$, if the unperturbed frequencies associated with each $\psi_{j}$ depend only on that phase, namely $\omega_{j}=\omega_{j}(\psi_{j},\ \mathbf{I})$, and if each $\psi_{j}$ is assumed to have a period of $2\pi$, then the adiabatic approximation of $f(\boldsymbol{\psi},\ \mathbf{I})$ can be evaluated as~\cite{PhysRevD.83.044037}
\begin{equation}
    \expval{f}_{\mathbf{I}}=\frac{\int_{0}^{2\pi}\mathrm{d}^{N}\boldsymbol{\psi}\prod_{j=1}^{N}\frac{1}{\omega_{j}(\psi_{j},\ I)}f(\boldsymbol{\psi},\ \mathbf{I})}{\prod_{j=1}^{N}\int_{0}^{2\pi}\mathrm{d}\psi_{j}\frac{1}{\omega_{j}(\psi_{j},\ I)}},
    \label{eq: average integral}
\end{equation}
where the numerator on the right-hand side denotes the $N$-dimensional integral over the phase set $\boldsymbol{\psi}$. It has been noted in Ref.~\cite{PhysRevD.83.044037} that for more general systems in which $\omega_{j}=\omega_{j}(\boldsymbol{\psi},\ \mathbf{I})$, the adiabatic approximation may not take the form of Eq.~\eqref{eq: average integral}.
In Schwarzschild spacetime, both frequencies in Eqs.~\eqref{eq:dpsidtau and dchidtau}
solely depend on phase $\psi_{r}$. This corresponds to a special case in which the adiabatic average can still be evaluated in a form analogous to Eq.~\eqref{eq: average integral} as 
\begin{equation}
    \expval{f}_{\mathbf{I}}=\frac{\int_{0}^{2\pi}d\psi_{\theta}\int_{0}^{2\pi}d\psi_{r}\frac{1}{\omega_{r}(\psi_{r},\ \mathbf{I})}f(\psi_{r},\ \psi_{\theta},\ \mathbf{I})}{2\pi\int_{0}^{2\pi}d\psi_{r}\frac{1}{\omega_{r}(\psi_{r},\ \mathbf{I})}},
    \label{eq:average integral1}
\end{equation}
note that the integration is carried out over the periodic domains of both $\psi_{r}$ and $\psi_{\theta}$, while only the $\omega_{r}$ appears in the denominator of the integrand. The equivalent form adopted in this work is shown in Appendix~\ref{appendix:average Integral2}. The adiabatic EOM is obtained by applying the average above to both sides of Eq.~\eqref{eq:parameter evolution ode}, 
\begin{equation}
    \dv{\expval{I_{0}}}{\tau}=\expval{C^{(I_{0})}_rf^{r}+C^{(I_{0})}_{\theta}f^{\theta}+C^{(I_{0})}_{\phi}f^{\phi}}\ ,
    \label{eq:secular ode}
\end{equation}
where  $\expval{C^{(I_{0})}_if^i}$ on the right-hand side can be precomputed for the purpose of numerical efficiency.


\section{Results}
\label{sec:results}

\subsection{Orbital evolution of stars}\label{sec:aerodynamic}
To assess the reliability of the adiabatic approximation in tracking secular evolution of a star colliding with an accretion disk, we first 
compare adiabatic evolution with full orbital evolution with the aero-drag force~\eqref{eq:aero drag}: semimajor axis $a(t)\equiv p/(1-e^2)$, the eccentricity $e(t)$ and the inclination angle $\iota(t)$.  

\begin{figure}
    \centering
    \includegraphics[scale=0.6]{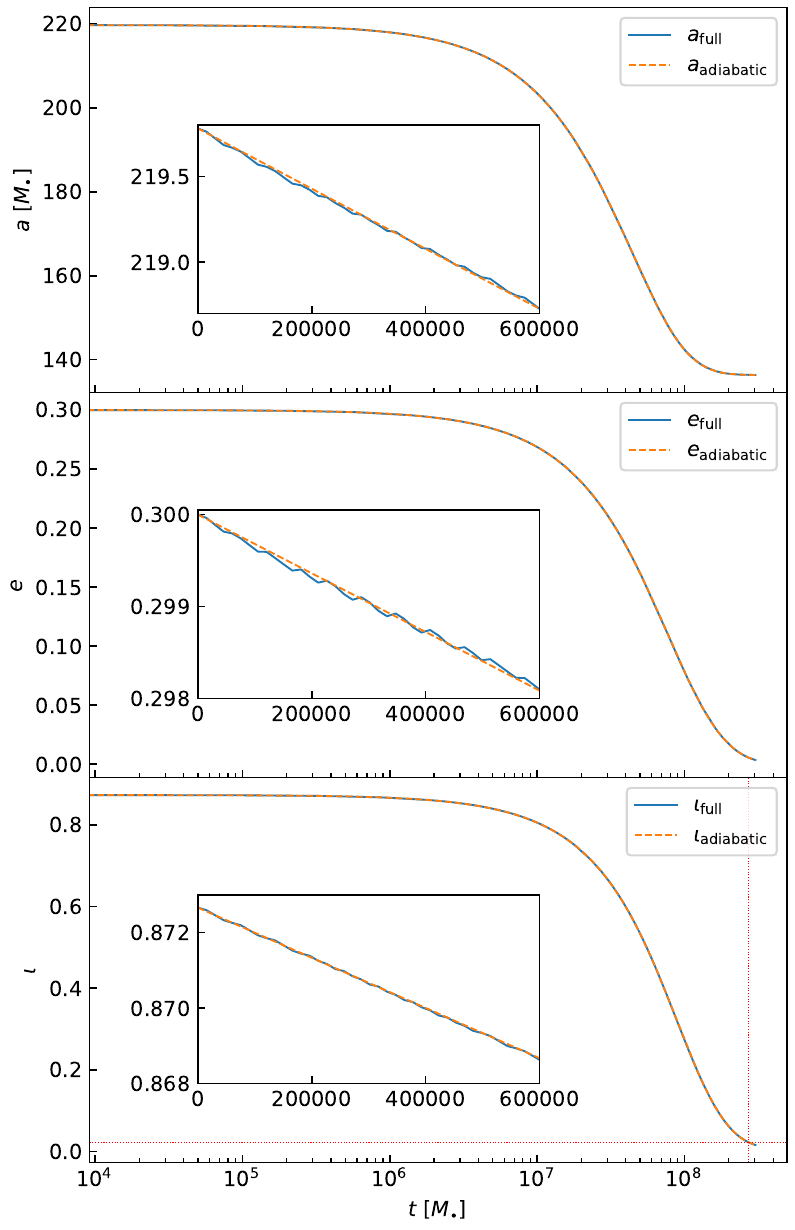}
    \caption{Evolution of the semimajor axis $a$, eccentricity $e$, and orbital inclination angle $\iota$ of a star. Solid lines labeled ``full" are obtained from full EOM \eqref{eq: forced geodesic}, while dashed lines labeled ``adiabatic" show the secular evolution results obtained by evolving  Eq.~\eqref{eq:secular ode}. The enlarged panels display the variations of $a,\ e,\ \iota$ over the relatively short initial time interval $(0,\ 6\times 10^5\ M_{\bullet})$. In the bottom panel, the intersection point of the two red dotted lines indicates the critical inclination $\iota_{\rm crit}=3H_{\rm disk}/p_{\rm ini}$ and the corresponding time $t_{\rm cap}=2.7\times 10^{8}\ M_\bullet$.}
    \label{fig:aerodynamic p 200 e 0.3 z1 40 deg three evo}
\end{figure}

As an example, we choose $\gamma_{0}=2\times 10^{-6}\ M_{\bullet}^{-1}$,  the initial conditions of the orbital parameters $p_{\rm ini}=200\ M_{\bullet},\ e_{\rm ini}=0.3,\ \iota_{\rm ini}=50^{\circ}$, and the initial phases  $\psi_{0\ \rm ini}=0$ and $\chi_{0\ \rm ini}=\pi/3$. The disk height to the midplane is set to $H_{\rm disk}=1.5\ M_{\bullet}$. As shown in Fig.~\ref{fig:aerodynamic p 200 e 0.3 z1 40 deg three evo}, the adiabatic approximation works well in tracking secular evolution of the stellar orbit.
The semimajor axis $a$ decays over time due to energy loss caused by star-disk collisions. Note that $a$ evolves slowly near the final stage when the star is close to being captured by the disk $\iota \rightarrow H_{\rm disk}/r$, because their relative speed and the drag force are greatly reduced. The eccentricity $e$ also decays with time and the orbit becomes
nearly circular with a final value $e_{\rm f}= 3.6\times 10^{-3}$ at the final stage of the evolution. Our numerical solution shows no secular evolution of the “initial” phases,  $\expval{\dot \psi_0} = \expval{\dot \chi_0} =0$, consistent with the expectation for the dissipative perturbing force as considered in this work~\cite{PhysRevD.83.044037}.

\begin{figure}
    \centering
    \includegraphics[scale=0.55]{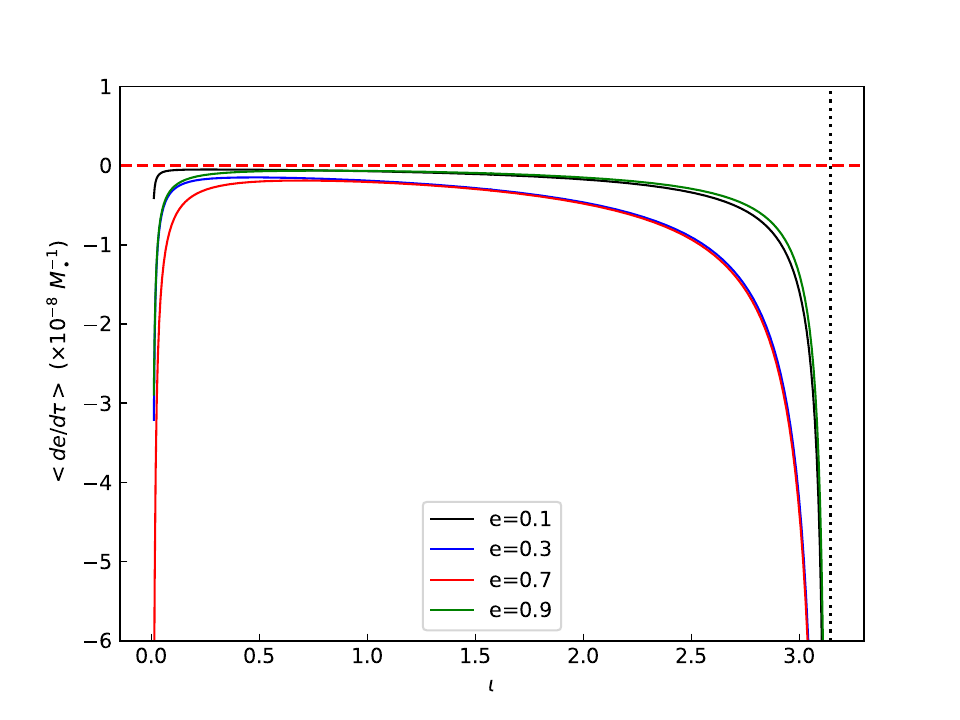}
    \caption{Comparison of $\expval{\mathrm{d}e/\mathrm{d}\tau}$ variations with orbital inclination angle $\iota$ for different $e$, with $p=300\ M_{\bullet}$ assuming the aero-drag force model. The vertical black dotted line marks $\iota=\pi\ (180^{\circ})$.}
    \label{fig:edot value aero}
\end{figure}

For the purpose of illustrating the eccentricity evolution driven by the aero-drag force, we compute the eccentricity evolution rate $\expval{\mathrm{d}e/\mathrm{d}\tau}$ as a function of the orbital inclination angle $\iota$, fixing the semilatus rectum at $p = 300\ M_{\bullet}$ and considering representative eccentricity values $e = 0.1,\ 0.3,\ 0.7,\ 0.9$, as shown in Fig.~\ref{fig:edot value aero}. It is evident that star-disk collisions tend to circularize the star orbit for all inclination angles and orbital eccentricities.


\begin{figure}
    \centering
    \includegraphics[scale=0.5]{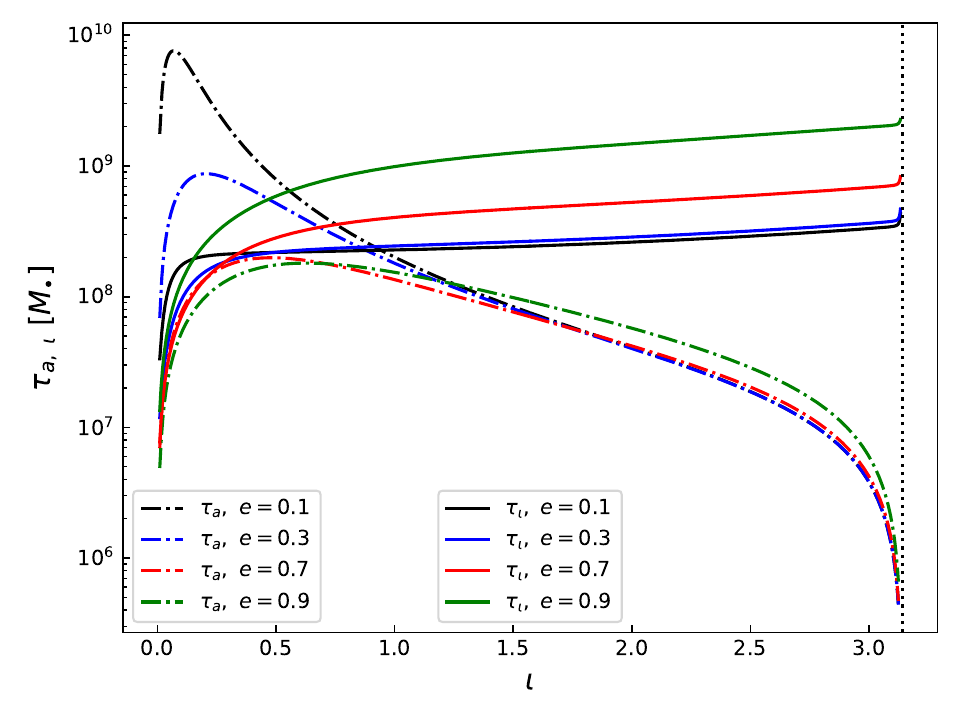}
    \caption{Comparison of estimated shrink timescale with $\abs{a/\expval{\dv{a}{\tau}}}$ (solid dotted line) and capture timescale with $\abs{\iota/\expval{\dv{\iota}{\tau}}}$ (solid line) with orbital inclination angle $\iota$ for different $e$, fixing $p=300\ M_{\bullet}$, using the aero-drag force model for stars. The vertical black dotted line marks $\iota=\pi\ (180^{\circ})$.} 
    \label{fig: timescale aero}
\end{figure}

To understand the final fate of a star colliding with a long-lived disk, we compare the disk-capture timescale $\tau_{\iota} := \abs{\iota/\expval{\dv{\iota}{\tau}}}$ and 
the orbital decay timescale $\tau_{a}:=\abs{a/\expval{\dv{a}{\tau}}}$ associated with the shrinkage of the semimajor axis $a$, where
\begin{equation}
    \expval{\dv{a}{\tau}}=\expval{\frac{1}{1-e^2}\dv{p}{\tau}+\frac{2ep}{(1-e^2)^2}\dv{e}{\tau}}.
\end{equation} 
As shown in Fig.~\ref{fig: timescale aero}, the capture timescale $\tau_{\iota}$ increases with increasing $\iota$, whereas the orbital decay timescale $\tau_{a}$ decreases due to the larger relative velocity between the star and the disk, which enhances the effect of the aero-drag force. As $\iota\rightarrow 0$, the timescale $\tau_{a}$ also decreases, however,  the star-disk collision model is no longer appropriate, since the orbit of the star becomes nearly embedded within the disk.

The comparison between $\tau_{\iota}$ and $\tau_{a}$ further shows that, for sufficiently large initial inclination angles, the orbital-decay timescale becomes shorter than the alignment timescale. This shows that a star on a high-inclination orbit will finally be disrupted by the  central SMBH before  being captured by the disk. 
On the other hand,  a star on a low-inclination orbit will be captured by the disk  before getting disrupted by the central SMBH.
However, we note that in the actual orbital evolution, the semimajor axis and eccentricity evolve simultaneously, which can modify the rates of change of the parameters. Consequently, such estimates based solely on the initial values may not fully coincide with the true evolutionary behavior. The above calculations also show that $\expval{\mathrm{d}{\iota}/\mathrm{d}{\tau}}<0$ and $\expval{\mathrm{d}{a}/\mathrm{d}{\tau}}<0$ over the entire range of $\iota$, consistent with the expectation and the actual evolution results found in this work.


\par

\begin{figure}
    \centering
    \includegraphics[scale=0.6]{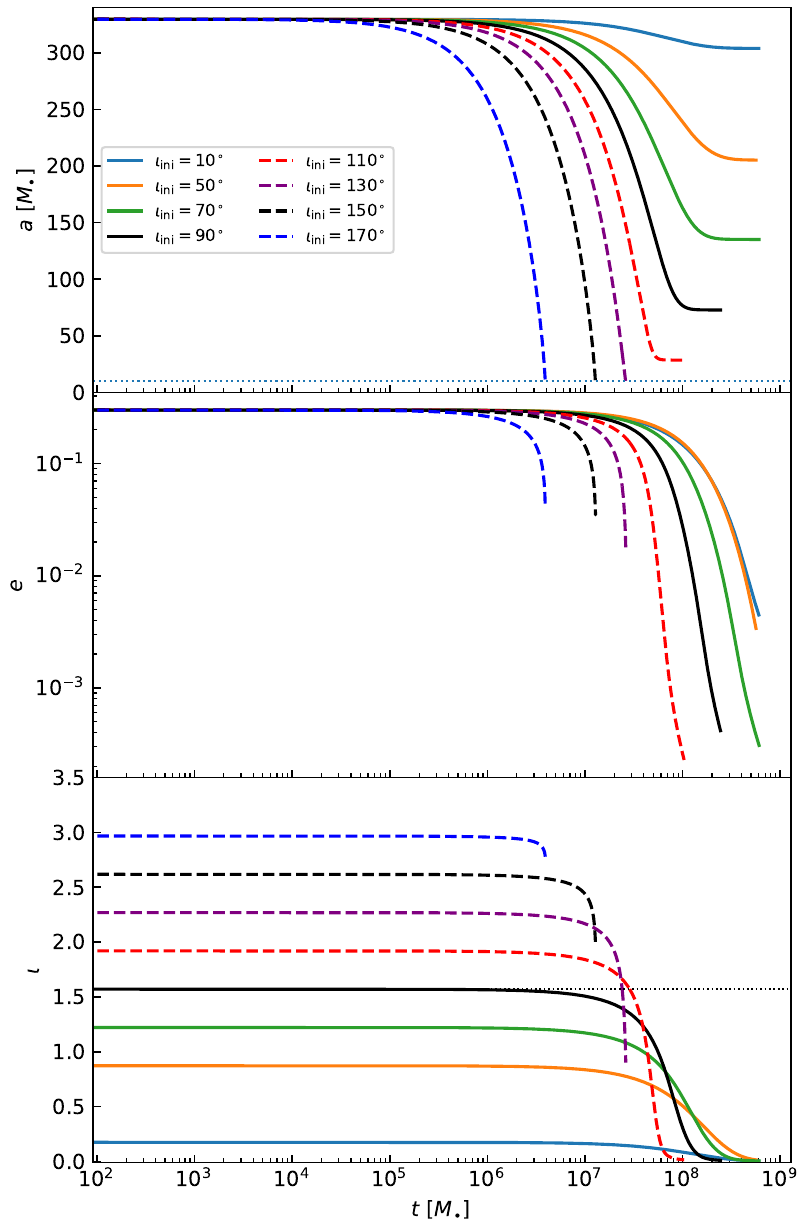}
    \caption{Evolution of the semimajor axis $a$, eccentricity $e$, and orbital inclination angle $\iota$ using the aero-drag force model with a constant damping coefficient $\gamma_{0}=2\times 10^{-6}\ M_{\bullet}^{-1}$. The initial conditions are $p_{\rm ini}=300\ M_{\bullet}$, and $\ e_{\rm ini}=0.3$. The prograde cases $(\iota_{\rm ini}<90^{\circ})$ are shown with solid lines, while the retrograde cases $(\iota_{\rm ini}>90^{\rm \circ})$ are shown with dashed lines. The blue dotted horizontal line at $a=10\ M_{\bullet}$ in the top panel marks the threshold below which the orbit is considered to have shrunk in this work. The black dotted horizontal line in the bottom panel marks $\iota=\pi/2$.}
    \label{fig:aero e03 three evo}
\end{figure}

After establishing the evolution behavior and the associated timescales for the orbital parameters $a,\ e$ and $\iota$ under the aero-drag force, we now present several representative numerical examples of the orbital evolution computed within the adiabatic approximation.
We consider orbits with an initial semilatus rectum $p_{\rm ini}=300\ M_{\bullet}$ and compute the evolution with different initial inclination angles, focusing on initial eccentricities of $e_{\rm ini}=0.3,\ 0.7$, under the aero-drag force model with $\gamma_{0}=2\times 10^{-6}\ M_{\bullet}^{-1}$.
The evolution of $a,\ e,\ \iota$ in the case of $e_{\rm ini}=0.3$ with different initial inclination angles under the aero-drag force model is shown in the three panels of Fig.~\ref{fig:aero e03 three evo}. 

As shown in the top and bottom panels of Fig.~\ref{fig:aero e03 three evo}, for the cases with $\iota_{\rm ini}\geq 130^{\circ}$, the orbits shrink, with the semimajor axis $a$ decreasing to below $10\ M_{\bullet}$, before alignment with the disk plane is achieved. For the cases with $\iota_{\rm ini}\leq 110^{\circ}$, the star successfully aligns with the disk plane and does not undergo significant shrinkage. The middle panel of Fig.~\ref{fig:aero e03 three evo} shows that the aero-drag force circularizes orbits with different inclination angles, regardless of the initial inclination.
We also find that stars on low-inclination orbits require longer capture times. This is primarily because, for low-inclination orbits, the relative velocity between the SMO and the disk material is small. According to Eq.~\eqref{eq:aero drag}, $\abs{\boldsymbol{F}_{\rm aero}}\propto \abs{\boldsymbol{v}_{\rm rel}}^{2}$, leading to a correspondingly slower orbital evolution. In addition, during the alignment process, the semimajor axis $a$ gradually decreases, which increases $\abs{\boldsymbol{v}_{\rm rel}}$ and accelerates the subsequent evolution. This effect further explains why low-inclination stars experience longer capture times: although high-inclination orbits require larger changes in inclination, they initially evolve more rapidly due to their larger relative velocities, and the accompanying shrinkage of $a$ during alignment further enhances this evolution, resulting in shorter capture times overall.

\begin{figure}
    \centering
    \includegraphics[scale=0.6]{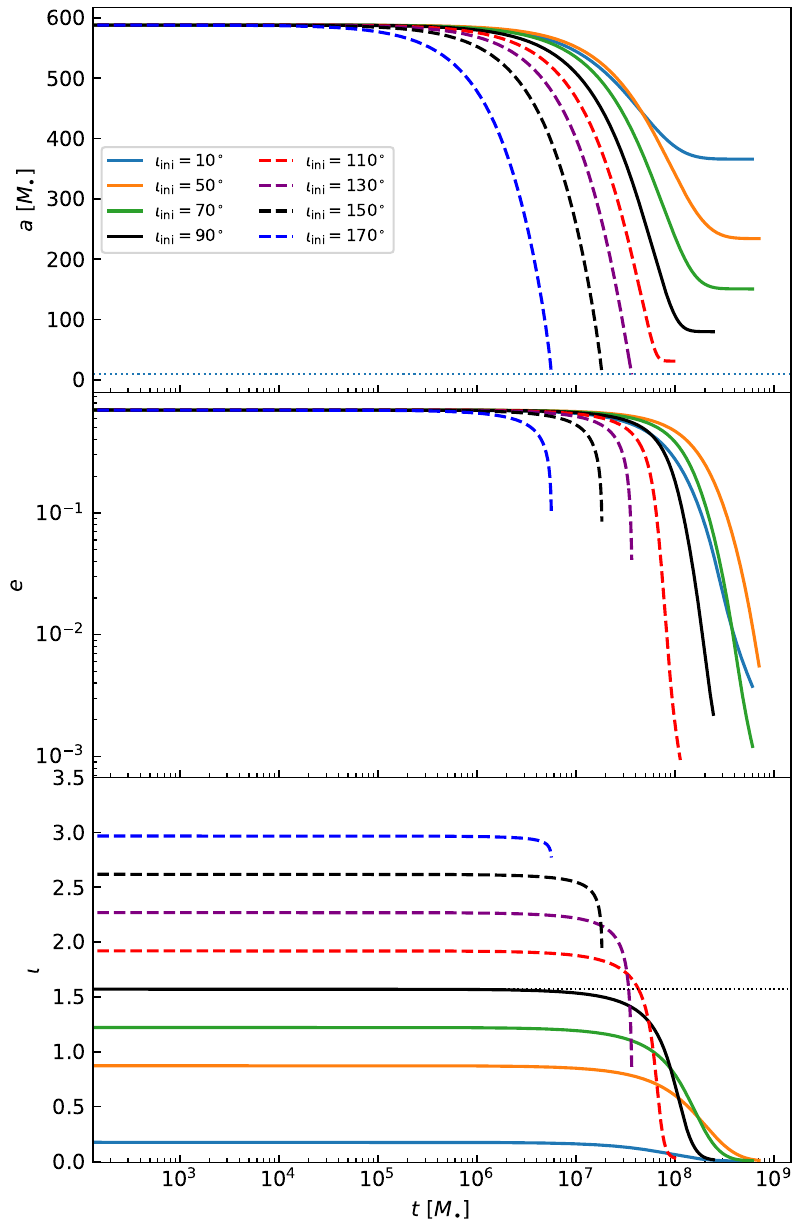}
    \caption{Same as Fig.~\ref{fig:aero e03 three evo} except for $e_{\rm ini}=0.7$.}
    \label{fig:aero e07 three evo}
\end{figure}

The evolution of $a,\ e,\ \iota$ in the case of $e_{\rm ini}=0.7$ with the aero-drag force model is shown in the three panels of Fig.~\ref{fig:aero e07 three evo}, with all other conditions identical to those for $e_{\rm ini}=0.3$. Overall, changing the initial eccentricity does not significantly affect the evolution of the orbital parameters.
\subsection{Orbital evolution of sBHs}
\begin{figure}
    \centering
    \includegraphics[scale=0.55]{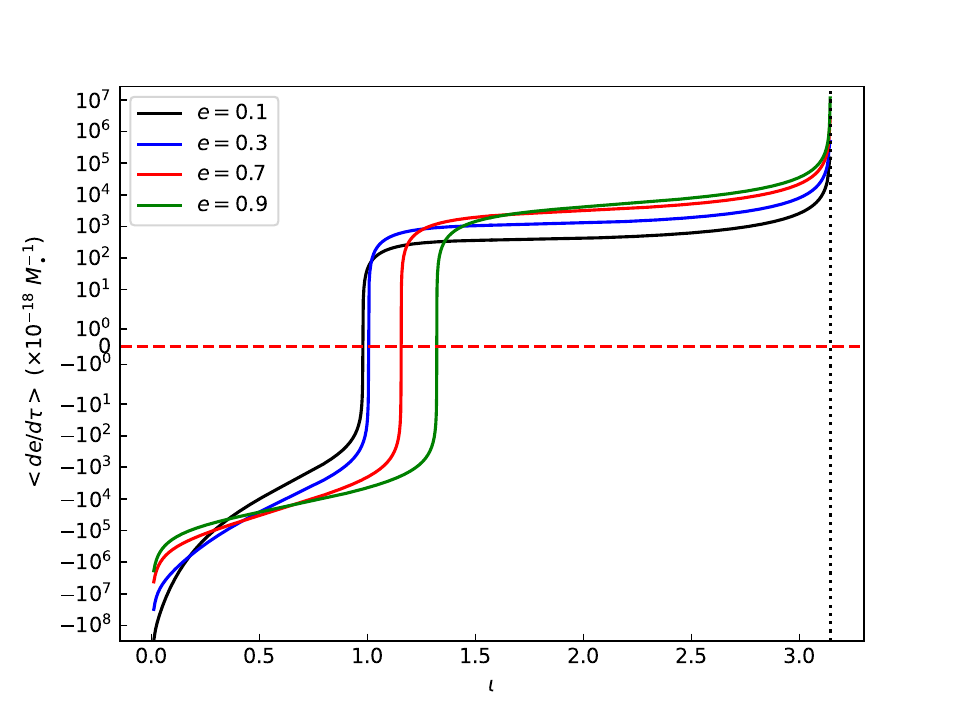}
    \caption{Comparison of $\expval{\dv{e}{\tau}}$ variations with orbital inclination angle $\iota$ for different $e$, with $p=300\ M_{\bullet}$, with the dynamical friction model for the sBH-SMO cases. The vertical black dotted line marks $\iota=\pi\ (180^{\circ})$.}
    \label{fig:edot value sbh}
\end{figure}

In this Subsection, we consider  sBH-disk collisions, where the perturbing force is dominated by dynamical friction~\eqref{eq:Fdyn}. The change rate of the eccentricity $e$ as a function of the orbital inclination angle $\iota$ is computed using the same method and initial conditions as in Fig.~\ref{fig:edot value aero}, except that the dynamical-friction model with $\gamma_{0}=4\times10^{-13}\ M_{\bullet}^{-1}$ is adopted. The results are shown in Fig.~\ref{fig:edot value sbh}. 
In contrast to the aero-drag force model, we find dynamical friction can increase the orbital eccentricity $e$
if the orbiter is highly misaligned  with $\iota$ in the range of $(\iota_{\rm crit}, 180^\circ)$, where $\iota_{\rm crit} \approx 60^\circ - 90^\circ$ for the parameter range considered in this work. Similar results were also found in previous studies \cite{Wang2024, spieksma2025gripdiskdraggingcompanion}, which motivated some recent discussions on eccentric wet EMRIs. As we will show later in this Section, EMRIs that are captured by disks are circularized due to the eccentricity damping when $\iota < \iota_{\rm crit}$.

\begin{figure}
    \centering
    \includegraphics[scale=0.5]{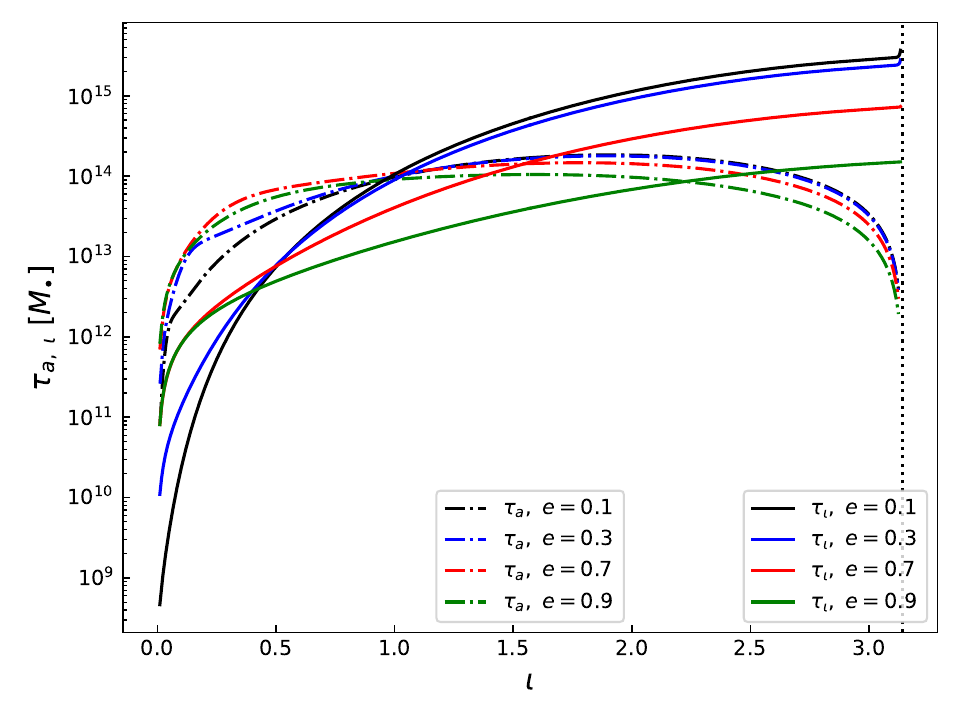}
    \caption{Comparison of estimated shrink timescale with $\tau_{a}=\abs{a/\expval{\dv{a}{\tau}}}$ (solid dotted line) and capture timescale with $\tau_{\iota}=\abs{\iota/\expval{\dv{\iota}{\tau}}}$ (solid line) with orbital inclination angle $\iota$ for different $e$, fixing $p=300\ M_{\bullet}$, using the dynamical friction force model for the sBH-SMO cases. The vertical black dotted line marks $\iota=\pi\ (180^{\circ})$.}
    \label{fig: timescale sbh}
\end{figure}

As discussed in Sec.~\ref{sec:aerodynamic}, we also calculate two timescales $\tau_{a}$ and $\tau_{\iota}$ for the sBH orbit, as shown in Fig.~\ref{fig: timescale sbh}, with the same set of initial conditions as in Fig.~\ref{fig:edot value sbh}. The calculations also yield $\expval{\mathrm{d}a/\mathrm{d}\tau}<0$ and $\expval{\mathrm{d}\iota/\mathrm{d}\tau}<0$ for all inclination angles, showing that the disk is aligning the orbiter. 
From Fig.~\ref{fig: timescale sbh}, for the sBH-SMO cases, large initial orbital inclination angles result in $\tau_{a}<\tau_{\iota}$, consistent with the behavior found in the stellar-SMO cases, while $\tau_{\iota}$ increases monotonically with $\iota$, $\tau_{a}$ initially increases with $\iota$ because the larger relative velocity between the sBH and the disk material reduces the effect of the dynamical friction according to Eqs.~\eqref{eq:Fdyn} and~\eqref{eq:dynamical friction gamma}, in contrast to the behavior under the aero-drag force model. It can also be found that $\tau_{a}$ decreases as $\iota\rightarrow\pi$, which can be interpreted as the strengthened interaction resulting from the sBH orbit gradually becoming embedded in the disk plane.
\par
\begin{figure}
    \centering
    \includegraphics[scale=0.6]{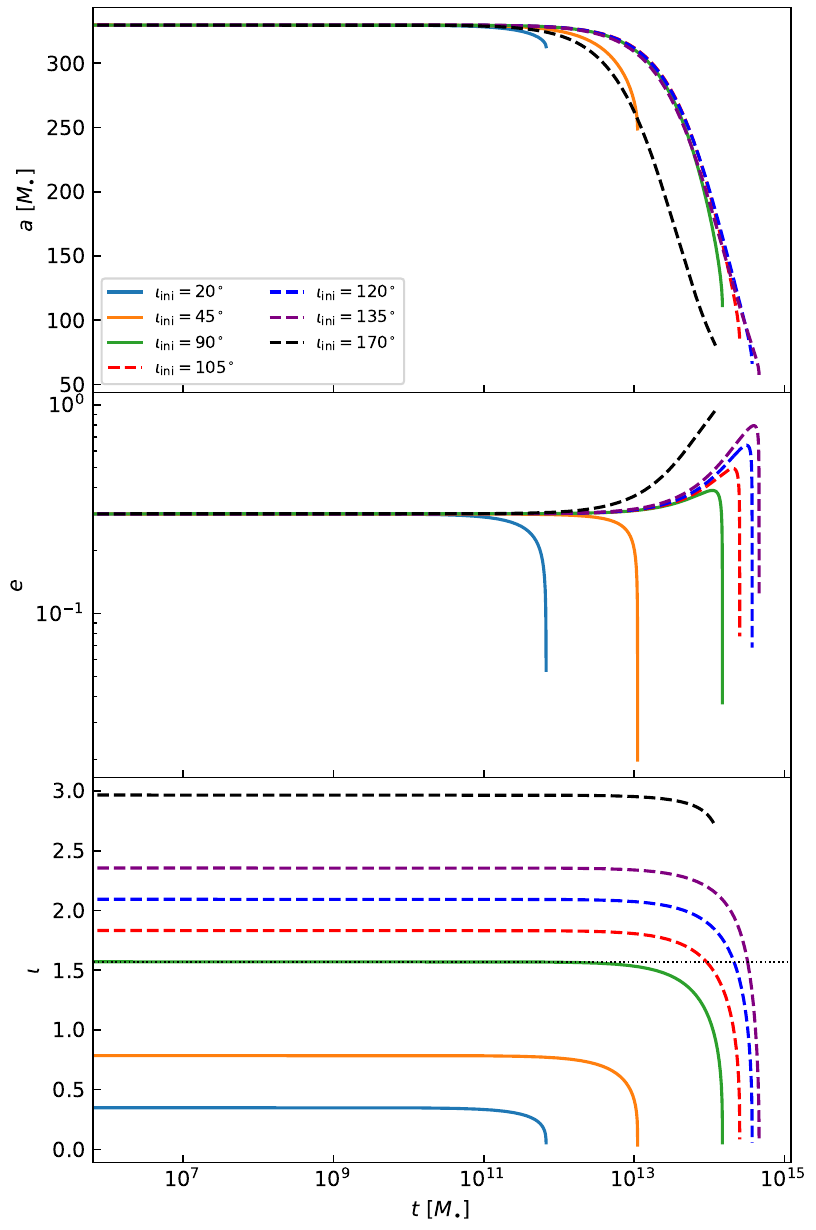}
    \caption{Same as Fig.~\ref{fig:aero e03 three evo} except for using the dynamical friction model with the constant part of damping coefficient $\gamma_{0}=4\times 10^{-13}\ M_{\bullet}^{-1}$. }
    \label{fig:friction e03 three evo}
\end{figure}

To illustrate the evolution behavior and timescale hierarchy discussed above, we now present representative numerical examples of the orbital evolution under the dynamical friction, computed within the adiabatic approximation. We consider orbits with an initial semilatus rectum $p_{\rm ini}=300\ M_{\bullet}$ and compute the evolution with different initial inclination angles, focusing on initial eccentricities of $e_{\rm ini}=0.3,\ 0.7$, under the dynamical-friction model with $\gamma_{0}=4\times 10^{-13}\ M_{\bullet}^{-1}$.
The evolution of $a,\ e,\ \iota$ for $e_{\rm ini}=0.3$ under the dynamical friction model is shown in the three panels of Fig.~\ref{fig:friction e03 three evo}. In this work, for dynamical friction model, the calculation is terminated once the orbit approaches the near-aligned configuration with the disk plane, as during the alignment process, the relative velocity between the sBH and the disk  gradually decreases, the change rate of the orbital parameters becomes sufficiently large and potentially invalidates the adiabatic approximation.

For cases with $\iota_{\rm ini}=20^{\circ}-135^{\circ}$, near-complete alignment of the orbits occurs before significant orbital shrinkage. In contrast, for the higher inclination case with $\iota_{\rm ini}=170^{\circ}$, although the semimajor axis $a$ is still several times that of $10\ M_{\bullet}$, the calculation is terminated before the condition $p-6-2e>0$ is violated~\cite{PhysRevD.50.3816}, which marks the transition from a stable to an unstable orbit and the inevitable infall of the sBH into the central SMBH.\par
As shown in the middle panel of Fig.~\ref{fig:friction e03 three evo}, for all retrograde cases, the eccentricity is excited (see Fig.~\ref{fig:edot value sbh}). For prograde cases, exemplified by $\iota_{\rm ini}=90^{\circ}$, the eccentricity is also initially excited. As $\iota$ decreases, a transition occurs from $\expval{\mathrm{d}e/\mathrm{d}\tau}>0$ to $\expval{\mathrm{d}e/\mathrm{d}\tau}<0$, and the eccentricity eventually declines. For smaller initial inclination angles, such as $\iota_{\rm ini}=20^{\circ},\ 45^{\circ}$, the eccentricity decreases from the outset.\par
\begin{figure}
    \centering
    \includegraphics[scale=0.6]{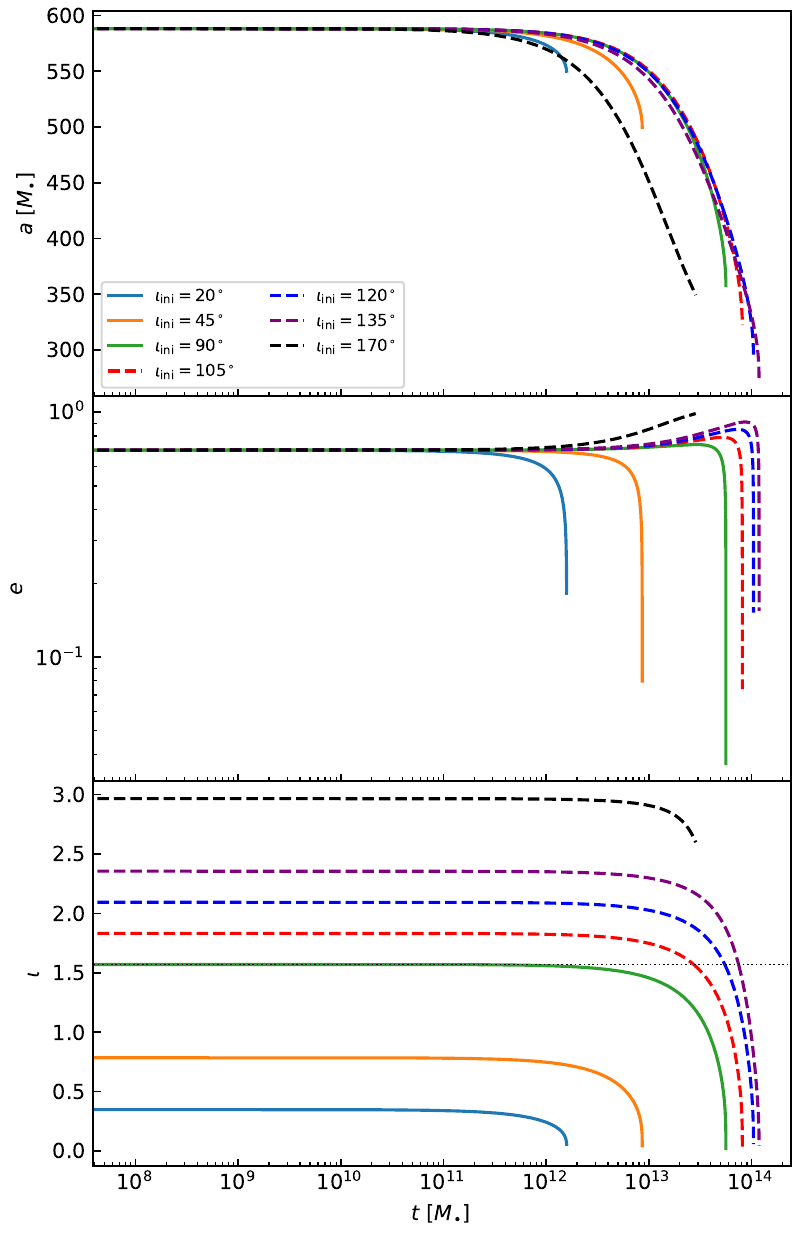}
    \caption{Same as Fig.~\ref{fig:friction e03 three evo} except for $e_{\rm ini}=0.7$.}
    \label{fig:friction e07 three evo}
\end{figure}

The evolution of $a,\ e,\ \iota$ for $e_{\rm ini}=0.7$ with the dynamical friction model is shown in the three panels of Fig.~\ref{fig:friction e07 three evo}, with all other conditions identical to those for $e_{\rm ini}=0.3$. Overall, the evolution of the orbital parameters shows no significant differences compared to the $e_{\rm ini}=0.3$ case.

\subsection{Capture timescales of AGN disks}\label{sec:capture timescales}
As shown in the preceding Subsections, collisions with the disk drive a secular decrease in the orbital inclination angle $\iota$, which may ultimately lead to the SMO being captured by the disk. In this Subsection, we estimate characteristic timescales for SMOs to be captured by the disk in EMRI systems over a wide range of orbital scales. 

We first analyze the dependence of the capture timescale $t_{\rm cap}$ on relevant quantities of the EMRI+disk system.
For clarity, starting from the aero-drag force model in Eq.~\eqref{eq:aero-drag force gamma}, we note that the coefficients $C_{z_1}(\theta),\ C_{z_{1}}(\phi)$ given in Appendix~\ref{appendix:coefficients} scale as $p^{3/2}$, while the $\theta,\ \phi$ acceleration components in Appendix~\ref{appendix:form of acceleration} scale as
\begin{equation}
    \gamma_{0}\ \abs{\boldsymbol{v}_{\rm rel}}\ p^{-3/2}\sim \gamma_{0}\ p^{-1/2}\ p^{-3/2}\sim \gamma_{0}\ p^{-2},
\end{equation}
 where $\abs{\boldsymbol{v}_{\rm rel}}$ can be approximated using the Keplerian relation that $\abs{\boldsymbol{v}_{\rm rel}}\sim p^{-1/2}$. Combining this with the crossing time of the SMO through the disk with the height $H_{\rm disk}$, which can be approximated using the Keplerian estimate $t_{\rm cross}\sim H_{\rm disk}\ v_{\rm SMO}^{-1}\sim H_{\rm disk}\ p^{1/2}$, and using $z_{1}=\sin\iota$, we find that the change in $\iota$ produced by each SMO-disk collision scales as
\begin{equation}
    \Delta\iota\sim\Delta z_{1}\sim p^{\frac{3}{2}}\cdot \gamma_{0}\ p^{-2}\cdot H_{\rm disk}\ p^{\frac{1}{2}}\sim \Sigma_{\rm g}R_{\star}^2m_{\star}^{-1}.
\end{equation}
Thus, the total number of collisions required for a given inclination angle $\iota$ to decrease from its initial value to nearly zero scales as
\begin{equation}
    N_{\rm cap}\sim \Sigma_{\rm g}^{-1}R_{\star}^{-2}m_{\star}.
\end{equation}
Since the orbital period of the SMO scales as $T_{\rm obt}\sim p^{3/2}$,  the capture timescale follows
\begin{equation}
    t_{\rm cap}\sim N_{\rm cap}\ T_{\rm obt}\sim \Sigma_{\rm g}^{-1}R_{\star}^{-2}m_{\star} p^{3/2}.
\end{equation}
The power law relations $t_{\rm cap}\sim p^{n}$ for the dynamical friction model can be derived in a similar manner. In both cases, the dependence of the damping coefficient on the relative velocity $\abs{\boldsymbol{v}_{\rm rel}}$ should be taken into account. The resulting power law relations for both interaction models are summarized in Table~\ref{tab:power tau p}.\par
\begin{figure}
    \centering
    \includegraphics[scale=0.55]{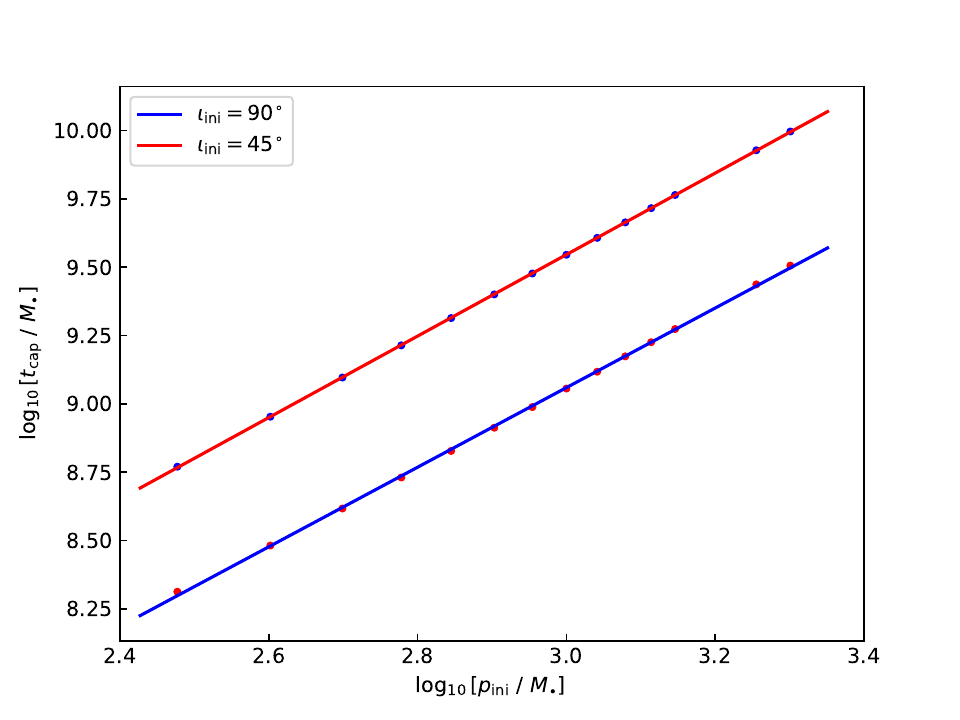}
    \caption{Dependence of the capture timescale on the initial semilatus rectum $p_{\rm ini}$ over the range $300\ M_{\bullet}$ to $2000\ M_{\bullet}$, computed using the aero-drag force model. }
    \label{fig:time_scale p list}
\end{figure}

As a numerical example, shown in Fig.~\ref{fig:time_scale p list}, we examine the dependence of the capture timescale $t_{\rm cap}$ on the initial semilatus rectum $p_{\rm ini}$ using the aero-drag force model with the damping coefficient $\gamma_{0}=2\times 10^{-6}\ M_{\bullet}^{-1}$. $p_{\rm ini}$ is sampled over in the range $(300,\ 2000)\ M_{\bullet}$, for two initial inclination angles, $\iota_{\rm ini}=45^{\circ}$ and $90^{\circ}$, while the initial eccentricity is fixed at $e_{\rm ini}=0.3$. For clarity, we define the capture timescale $t_{\rm cap}$ as the coordinate time at which the orbital inclination decreases from its initial value $\iota_{\rm ini}$ to a small threshold $\iota_{\rm crit}=1.5\times10^{-2}$.
The numerical results are well consistent with the expected scaling  $t_{\rm cap}\sim p^{3/2}$.

\begin{table}[htbp]
  \caption{Power scaling relations between $t_{\rm cap}$ and $p$ and other parameters}
  \label{tab:power tau p}
  \begin{tabular}{l|ccc}
    Model & $f^{\mu}$ & $N_{\rm cap}$ & $t_{\rm cap}$ \\
    \hline
    Aero-drag & $\sim \gamma_{0}\abs{\bm{v}_{\rm rel}}u^{\mu}_{\rm rel}$ & $\sim \Sigma_{\rm g}^{-1} R_{\star}^{-2}m_{\star}p^{0}$ & $\sim \Sigma_{\rm g}^{-1} R_{\star}^{-2}m_{\star}p^{\frac{3}{2}}$ \\
    Dynamical friction & $\sim \gamma_{0}\abs{\bm{v}_{\rm rel}}^{-3}u^{\mu}_{\rm rel}$ & $\sim\Sigma_{\rm g}^{-1} m_{\bullet}^{-1} p^{-2}$ & $\sim \Sigma_{\rm g}^{-1} m_{\bullet}^{-1}p^{-\frac{1}{2}}$ \\
  \end{tabular}
\end{table}
\label{sec:discussion}
\par
\begin{figure}[htbp]
    \centering
    \includegraphics[scale=0.55]{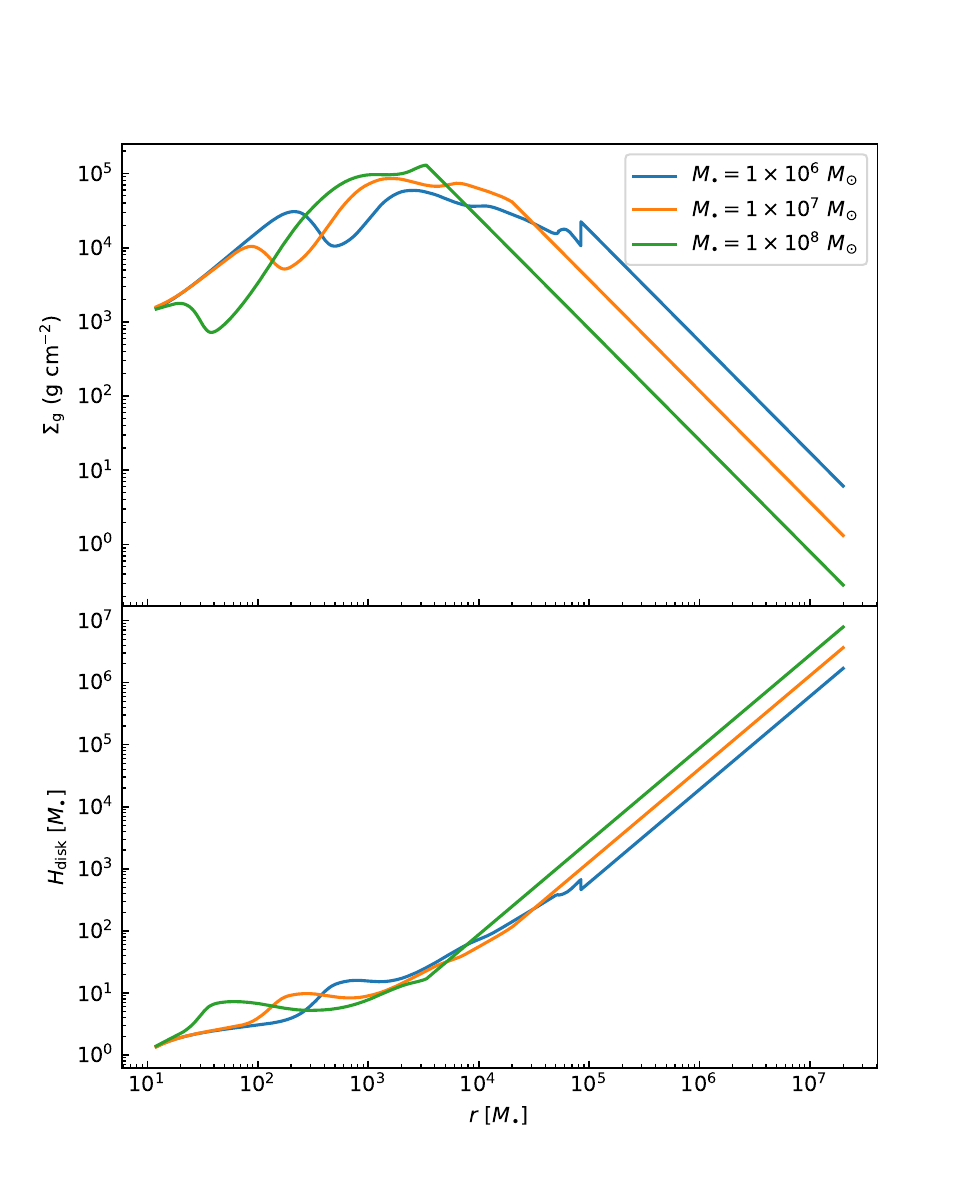}
    \caption{The surface density $\Sigma_{\rm g}$ and disk height $H_{\rm disk}$ distribution with the radial distance $r$ from the central SMBH with different masses, for the SG disk model, with the Eddington ratio $l_{e}=0.5$, viscosity parameter $\alpha=0.1$, hydrogen abundance $X=0.7$, and $\alpha-$disk case with $b=0$. }
    \label{fig:Sigmag Hdisk pagn}
\end{figure}

To assess whether misaligned EMRIs can be captured by the AGN disk within a certain timescale, we estimate the disk-capture timescale using scaling relations in Table~\ref{tab:power tau p}. The capture timescale is evaluated with $p_{\rm ini}=300\ M_{\bullet}$, and $e_{\rm ini}=0.3$, in combination with the Sirko and Goodman (SG) disk model~\cite{10.1046/j.1365-8711.2003.06431.x}.
For simplicity, we restrict our analysis to cases in which orbital alignment occurs prior to significant orbital shrinkage. 
We adopt the same conventions as in Ref.~\cite{Wang2024} for the SMO mass, taking $m_{\star}=m_{\bullet}=30\ M_{\odot}$, and for stellar SMOs, the stellar radius is set to $R_{\star}=R_{\odot}(m_{\star}/M_{\odot})^{\frac{3}{4}}$. 
The disk surface density and scale-height profiles are obtained using the \texttt{Sirko} module in the \texttt{PAGN} package ~\cite{Gangardt_2024} (see Fig.~\ref{fig:Sigmag Hdisk pagn}).

\begin{figure}[htbp]
    \centering
    \includegraphics[scale=0.55]{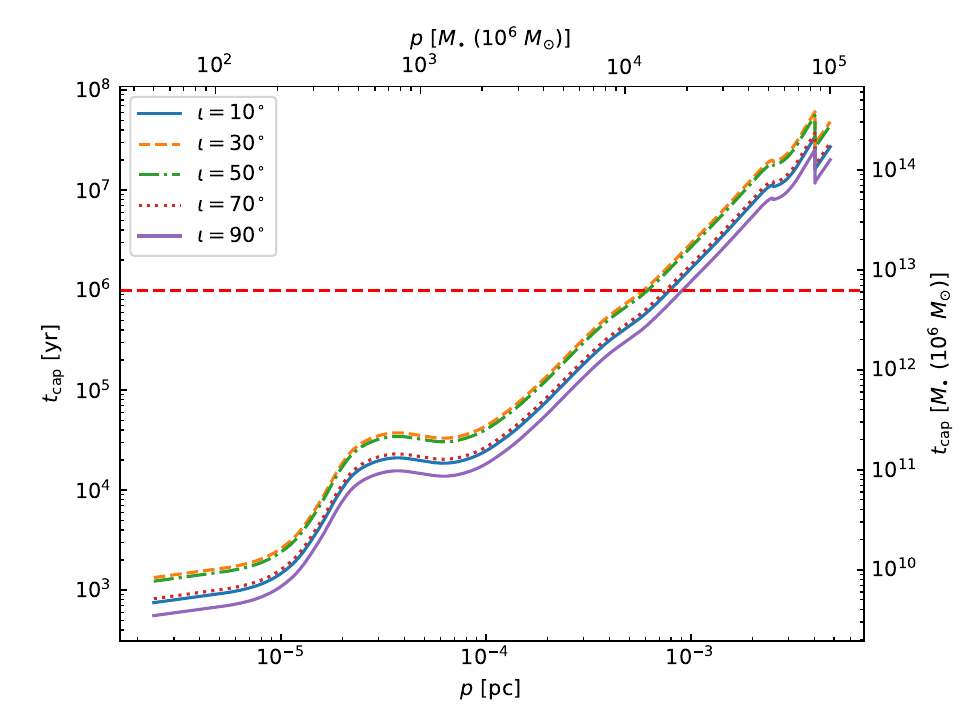}\\
    \includegraphics[scale=0.55]{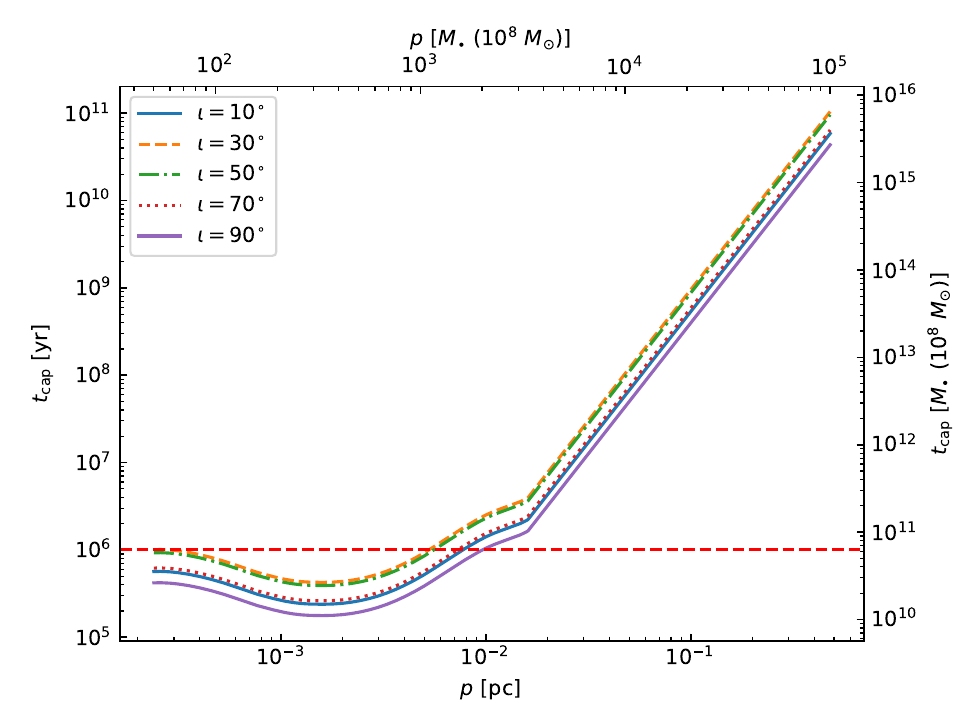} 
    \caption{Capture timescales estimated using the aero-drag force model for initial inclination angles $\iota_{\rm ini}=10^{\circ},\ 30^{\circ},\ 50^{\circ},\ 70^{\circ},\ 90^{\circ}$. The horizontal red dashed line indicates $t=10^{6}\ \mathrm{yr}$. The top and bottom panels correspond to SMBH masses of $M_{\bullet}=10^{6}\ M_{\odot}$ and $10^{8}\ M_{\odot}$, respectively. We note that the range of $p$ is truncated at $10^{5}M_{\bullet}$ to keep the stellar cases within the applicable regime of the aero-drag prescription.} 
    \label{fig:aero estimation 1e6 1e8}
\end{figure}

For the aero-drag force model, the results presented in Figs.~\ref{fig:aero e03 three evo} and~\ref{fig:aero e07 three evo} show that the capture timescale in prograde cases is longer than  in retrograde cases.
The estimates are carried out for initial inclination angles $\iota_{\rm ini}=10^{\circ},\ 30^{\circ},\ 50^{\circ},\ 70^{\circ},\ 90^{\circ}$ and for SMBH masses of $M_{\bullet}=10^6\ M_{\odot}, 10^{8}\ M_{\odot}$, with the results shown in two panels of Fig.~\ref{fig:aero estimation 1e6 1e8}, respectively.
For prograde cases under the aero-drag force model, the capture timescale exhibits only a weak dependence on the initial orbital inclination angle, as can be seen in Figs.~\ref{fig:aero e03 three evo} and \ref{fig:aero e07 three evo}. This behavior can be interpreted as follows. Larger orbital inclinations lead to higher relative velocities, which in turn produce larger inclination-change rates. As a result, once the inclination exceeds a certain threshold, the alignment timescale decreases.\par
\begin{figure}[htbp]
    \centering
    \includegraphics[scale=0.55]{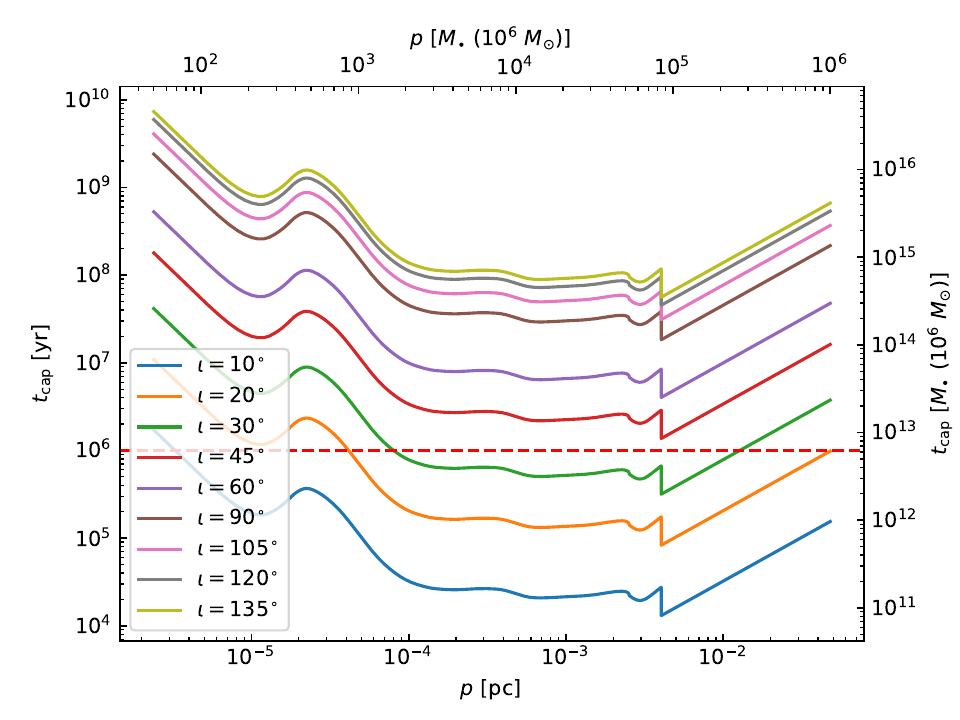}\\
     \includegraphics[scale=0.55]{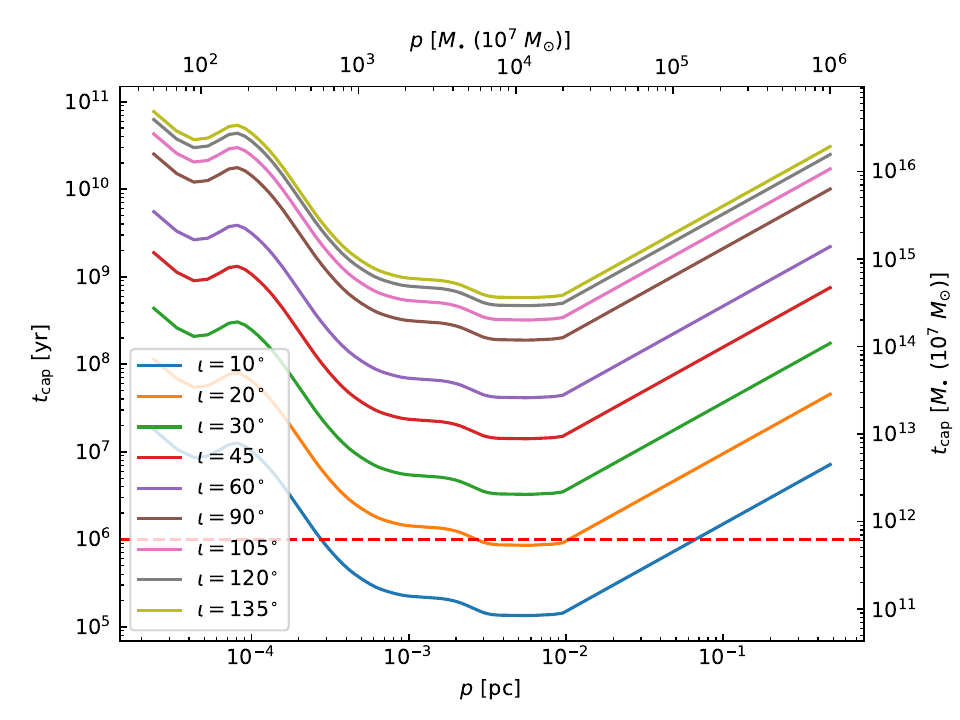}\\
      \includegraphics[scale=0.55]{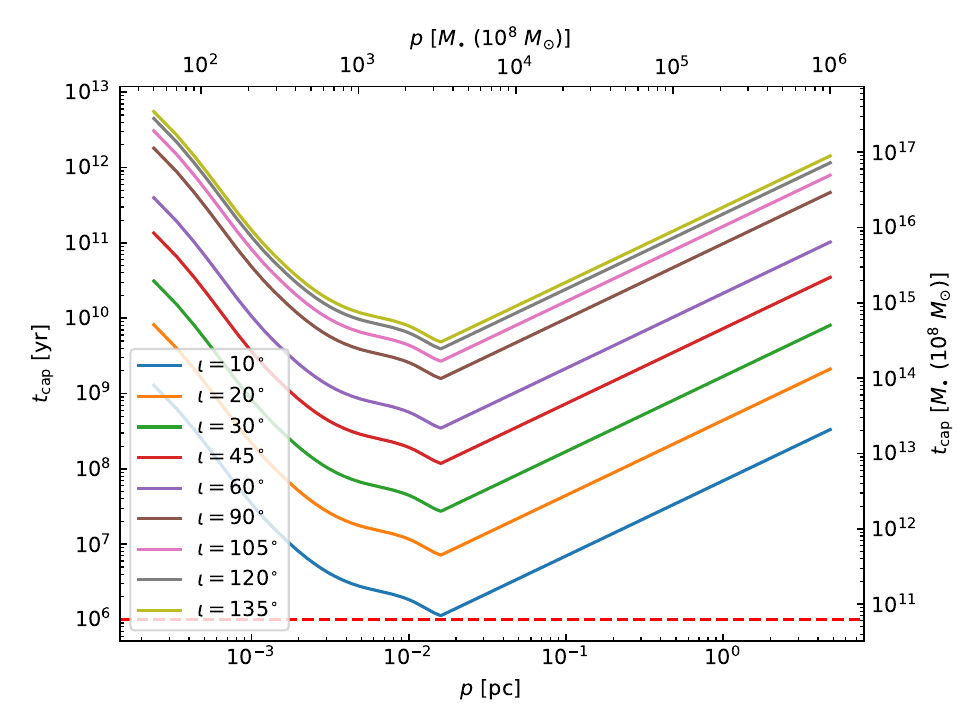}
    \caption{Capture timescales estimated using the dynamical friction model for initial inclination angles $\iota_{\rm ini}=10^{\circ},\ 20^{\circ},\ 30^{\circ},\ 45^{\circ},\ 60^{\circ},\ 90^{\circ},\ 105^{\circ},\ 120^{\circ},\ 135^{\circ}$. The horizontal red dashed line indicates $t=10^{6}\ \mathrm{yr}$. The top, middle and bottom panels correspond to SMBH masses of $M_{\bullet}=10^{6}\ M_{\odot},\ 10^{7}\ M_{\odot}$ and $10^{8}\ M_{\odot}$, respectively.}
    \label{fig:sbh estimation 1e6 1e7 1e8}
\end{figure}

For the dynamical friction model, the timescale estimates are evaluated for SMBH masses of $M_{\bullet}=10^6\ M_{\odot},\ 10^{7}\ M_{\odot}$ and $10^8\ M_{\odot}$, as shown in three panels of Fig.~\ref{fig:sbh estimation 1e6 1e7 1e8}, respectively.
In the dynamical friction model, the capture timescale increases significantly with orbital inclination, since larger relative velocities lead to smaller inclination-change rates.\par
For both models, the capture timescale is compared with a critical timescale of $10^{6}$ yr (1 Myr).
For the aero-drag force model, SMOs at distances of order $10^{-3}\ \rm pc$ can be captured within 1 Myr for $M_{\bullet}=10^6\ M_{\odot}$, whereas this distance increases to about $10^{-2}\ \rm pc$ for $M_{\bullet}=10^{8}\ M_{\odot}$.
For the dynamical friction model, the range of orbital inclination angles for which SMOs can be captured within 1 Myr decreases noticeably with increasing SMBH mass, primarily due to the disk surface-density profile and the scaling between the SMBH mass $M_{\bullet}$ and physical time. 

\subsection{Comparison with previous studies}\label{sec:comparison}
In previous studies~\cite{spieksma2025gripdiskdraggingcompanion,Wang2024,Rowan_2025,10.1093/mnras/staa3004,10.1093/mnras/stad1295,Secunda_2021,MacLeod:2019jxd,OConnor2020} of star-disk collisions in EMRIs, the orbital evolution of the SMO is typically modeled within a Newtonian framework. The secular changes are commonly estimated by summing the contributions from the two disk crossings within one orbital period and dividing by the orbital period. In this Section, we compare the evolution of the orbital parameters $a,\ e$ and $\iota$ for the stellar-SMO and sBH-SMO cases, obtained under the adiabatic approximation in Schwarzschild spacetime and shown in Figs.~\ref{fig:aero e03 three evo}–\ref{fig:friction e07 three evo}, with those reported in previous studies.\par

Our calculations broadly agree with the results of \cite{Wang2024}.
Consistent with \cite{spieksma2025gripdiskdraggingcompanion} and \cite{Wang2024}, we find that collisions with the disk always tend to align the SMO no matter what the initial orbital inclination $\iota_{\rm ini}$ is.
An opposite trend was claimed in \cite{10.1093/mnras/stad1295} for sBHs with large initial inclination angles. 
The authors cautioned that this peculiar behavior might reflect missing physics, such as the inconsistent assumption of circular orbits and the neglect of disk backreaction on the orbiting object.
Consistent with \cite{spieksma2025gripdiskdraggingcompanion} and \cite{Wang2024} , we find star-disk collisions always circularize the star orbit, while sBH-disk collisions excite the orbital eccentricity when the orbital inclination $\iota$ is large. But the temporary orbital eccentricity excitation does not lead to formation of eccentric wet EMRIs because the orbital eccentricity is largely damped when the inclination decreases and approaches zero when the orbiter is captured by the disk (Figs.~\ref{fig:friction e03 three evo}, \ref{fig:friction e07 three evo}). 

In \cite{Secunda_2021},  the authors found that the orbital eccentricity of embedded retrograde EMRIs can be efficiently excited to near unity by dynamical friction. This behavior is qualitatively consistent with the eccentricity growth observed in our high-inclination sBH cases, particularly for $\iota_{\rm ini}=170^{\circ}$ (see Figs.~\ref{fig:friction e03 three evo}, \ref{fig:friction e07 three evo}).
The finding has been used for argument of formation of eccentric wet EMRIs in AGN disks. In fact, as we have found in this work by tracking full orbital evolution of EMRIs,  AGN disks always tend to align EMRIs that are initially misaligned, and the captured EMRIs are effectively circularized though the orbital eccentricity may be temporarily excited during the capture process (Figs.~\ref{fig:friction e03 three evo}, \ref{fig:friction e07 three evo}).  As a result, there is little parameter space for forming eccentric wet retrograde EMRIs. 
\par

We also note that the orbits are not yet fully circularized within the portion of the evolution directly presented in this work. As shown in Figs.~\ref{fig:friction e03 three evo} and~\ref{fig:friction e07 three evo}, the sBH cases still retain a residual eccentricity of $e\lesssim  0.1$ near the capture stage, similar to the results in~\cite{Wang2024}. Nevertheless, the eccentricity decreases as the system approaches alignment, and further damping is expected in the subsequent near-coplanar and embedded phases, where density-wave torques are likely to dominate over impulsive SMO-disk collisions~\cite{9src-p7sp,duque2025extrememassratioinspiralsrelativisticaccretion,dyson2026spiraldensitywavestorque,g83s-jdld}. Reference~\cite{PhysRevD.111.084006} found that coplanar embedded EMRIs generally undergo efficient eccentricity damping in the subsonic regime relevant to the late-stage systems considered here, although the damping is weaker in the supersonic regime, with circularization timescale $t_e$ exceeding the shrinking timescale $t_a$ at $e\gtrsim0.85$. We therefore expect the residual eccentricity at capture stage to be further reduced during the subsequent embedded evolution.
\par
In addition to qualitative agreement, our calculations also show quantitative agreement with \cite{Wang2024} e.g., in the capture timescale of sBHs by AGN disks. 
For the sBHs, we adopt a constant damping coefficient $\gamma_{0}=4\times 10^{-13}\ M_{\bullet}^{-1}$, together with the disk scale height $H_{\rm disk}=1.5\ M_{\bullet}$. According to Eq.~\eqref{eq:dynamical friction gamma}, the remaining factors entering the damping coefficient can be combined into the dimensionless normalization $(\Sigma_{\rm g}/10^5\ \mathrm{g\cdot cm^{-2}})\cdot(m_{\bullet}/M_{\odot})\approx 86.58$. In Ref.~\cite{Wang2024}, sBHs were modeled with an initial semimajor axis $a_{\rm ini}=10^{-3}\ \mathrm{pc}$ and eccentricity $e_{\rm ini}=0.7$, which corresponds to an initial semilatus rectum $p_{\rm ini}\sim 10^2\ M_{\bullet}$, for a central SMBH of $M_{\bullet}=10^{8}\ M_{\odot}$.
Adopting the surface density profile implied by the $\alpha-$disk model shown in their work, and using the scaling relations listed in Table~\ref{tab:power tau p}, together with the numerical results obtained in this work (see Fig.~\ref{fig:friction e07 three evo}), we find that the resulting capture timescales are consistent at the order-of-magnitude level. We note that the capture timescales reported in Ref.~\cite{Wang2024} are systematically shorter than those obtained here, likely because their disk model implies an increasing surface density as the orbit shrinks, whereas a constant surface density assumption is adopted in this work; nevertheless, the overall scaling remains consistent.
\section{Conclusion}\label{sec:conclusion}
In this work, we study the orbital evolution of EMRI driven by interactions with accretion disks in a relativistic framework. We derive the secular evolution of the key orbital parameters, the semimajor axis $a$, eccentricity $e$, and orbital inclination angle $\iota$, by adopting the double-phase average formulation~\eqref{eq:average integral1}, which is suitable for adiabatic evolution over the proper time $\tau$.
Building on this relativistic framework, we develop a theoretical model for SMO-disk collisions under the thin-disk assumption and analyze the resulting orbital evolution under two representative interaction mechanisms: aero-drag and dynamical friction forces, which capture the essential features of star and sBH crossings.\par
Under both the aero-drag and dynamical friction forces, $a$ and $\iota$ decrease monotonically in both prograde and retrograde configurations, reflecting orbital-energy dissipation and the tendency toward angular-momentum alignment between the orbit and the disk. The eccentricity $e$ decreases monotonically under aero-drag force, leading to a steadily circularizing orbit, whereas under dynamical friction it can be excited at sufficiently large inclinations in some prograde cases and throughout the retrograde cases. In addition, based on the numerical results, we find that the orbit  becomes nearly circularized as the SMO is captured by the disk or as its orbit shrinks.\par
Based on the SG disk model and the scaling relations derived for different interaction mechanisms, together with the alignment times inferred from the representative cases computed in this work, we estimate the capture timescales for wider orbits around SMBHs of various masses and assess the feasibility of capturing SMOs within the timespan of 1 Myr.
We find only a small fraction of sBHs that are initially close to the SMBH and close to the disk
can be captured by the disk within the typical disk lifetime of active galactic nuclei. Therefore, two-body scatterings between sBHs and stars in the nuclear stellar cluster  play an essential role in randomly kicking EMRIs toward the disk and boosting the capture rate \cite{PhysRevD.105.083005}.

We have also derived a power-law dependence of the capture timescale on the initial semilatus rectum $p$. Specifically, we find $t_{\rm cap}\sim p^{3/2}$ for the aero-drag force model and $t_{\rm cap}\sim p^{-1/2}$ for the dynamical friction model. These scalings are obtained based on the evolution equation of the orbital inclination angle $\iota$ and the disk-crossing process. 
We then compare the orbital evolution and the capture timescales obtained in this work with those reported in several previous studies. We find that the orbital evolution behavior identified here is broadly consistent with most of the existing literature. Moreover, by applying the scaling relations listed in Table~\ref{tab:power tau p} together with more realistic disk surface density models, the estimated capture timescales can be consistent with previous results at the order-of-magnitude level. \par

\appendix
\section{Basic form of acceleration components}\label{appendix:form of acceleration}
With the basic acceleration form of Eq.~\eqref{eq:force form}, with $u^{\mu}_{\rm ZAMO}$ substituted with $u^{\mu}_{\rm disk}$, the $r,\ \theta,\ \phi$ components of the acceleration can be expressed as
\begin{equation}
\begin{aligned}
    f^{r}(\psi,\ \chi)&=-\gamma_{\rm gas}(\psi,\ \chi)u^{r}\\
    &=-\gamma_{\rm gas}(\psi,\ \chi)\sqrt{\frac{p-6-2e\cos(\psi-\psi_{0})}{p(p-3-e^2)}}e\sin(\psi-\psi_{0}),
\end{aligned}
\label{eq:ar}
\end{equation}
\begin{equation}
\begin{aligned}
    f^{\theta}(\psi,\ \chi)&=-\gamma_{\rm gas}(\psi,\ \chi)u^{\theta}\\
    &=-\gamma_{\rm gas}(\psi,\ \chi)\frac{[1+e\cos(\psi-\psi_{0})]^2}{p\sqrt{p-3-e^2}}\cdot\frac{z_{1}\sin(\chi-\chi_{0})}{\sqrt{1-z_{1}^{2}\cos^{2}(\chi-\chi_{0})}},
\end{aligned}
\label{eq:atheta}
\end{equation}
\begin{equation}
    \begin{aligned}
        f^{\phi}(\psi,\ \chi)&=-\gamma_{\rm gas}(\psi,\ \chi)\left(u^{\phi}+\frac{u^{\phi}_{\rm gas}}{u^{\rm gas}_{t}u^{t}+u^{\rm gas}_{\phi}u^{\phi}}\right)\\
        &=-\gamma_{\rm gas}(\psi,\ \chi)\left\{\frac{[1+e\cos(\psi-\psi_{0})]^{2}}{p\sqrt{p-3-e^2}}\cdot\frac{\sqrt{1-z_{1}^2}}{1-z_{1}^{2}\cos^{2}(\chi-\chi_{0})}\right.\\
        &\left.+\frac{\sqrt{p-3-e^2}}{p\sqrt{1-z_{1}^{2}}-\frac{p\sqrt{(p-2)^2-4e^2}}{\left[1+e\cos(\psi-\psi_{0})\right]^{\frac{3}{2}}}}\right\}.
    \end{aligned}
\label{eq:aphi}
\end{equation}

\section{Coefficients $C_{r,\theta,\phi}^{(I_0)}$}\label{appendix:coefficients}
The detailed expressions of the coefficients for $p,\ e,\ $ and $z_{1}$ in Eq.~\eqref{eq:parameter evolution ode} are
\begin{equation}
\begin{aligned}
     C^{(p)}_{r}&=-\frac{2(p-3-e^2)}{(p-6)^2-4e^2}\sqrt{\frac{e^2p^3[p-2e\cos(\psi-\psi_{0})-6]}{p-3-e^2}}\\
     &\cdot\sin(\psi-\psi_{0}),
\end{aligned}
\end{equation}
\begin{equation}
    \begin{aligned}
    C^{(p)}_{\theta}&=\frac{2p^2 z_{1}\sin(\chi-\chi_{0})\sqrt{p-3-e^2}}{[(p-6)^2-4e^2]\sqrt{1-z_{1}^{2}\cos(\chi-\chi_{0})^{2}}[1+e\cos(\psi-\psi_{0})]^2}\\
    &\cdot[18-9p+p^2-2e(-3+p)\cos(\psi-\psi_{0})-e^2(-6+p)\\
    &\cdot\cos^{2}(\psi-\psi_{0})+2e^{3}\cos^{3}(\psi-\psi_{0})],
\end{aligned}
\end{equation}
\begin{equation}
    \begin{aligned}
        C^{(p)}_{\phi}&=\frac{2p^2\sqrt{(p-3-e^2)(1-z_{1}^{2})}}{[(p-6)^{2}-4e^{2}][1+e\cos(\psi-\psi_{0})]^{2}}\\
        &\cdot[18-9p+p^{2}-2e(-3+p)\cos(\psi-\psi_{0})-e^{2}(-6+p)\\
        &\cdot\cos^{2}(\psi-\psi_{0})+2e^{3}\cos^{3}(\psi-\psi_{0})],
    \end{aligned}
\end{equation}
\begin{equation}
\begin{aligned}
     C^{(e)}_{r}&=\frac{(p-6-2e^2)\sin(\psi-\psi_{0})}{(p-6)^{2}-4e^2}\\
     &\cdot\sqrt{p(p-3-e^{2})[p-6-2e\cos(\psi-\psi_{0})]},
\end{aligned}
\end{equation}
\begin{equation}
    \begin{aligned}
        C^{(e)}_{\theta}&=\frac{pz_{1}\sin(\chi-\chi_{0})\sqrt{p-3-e^{2}}}{[(p-6)^{2}-4e^{2}]\sqrt{1-z_{1}^{2}\cos^{2}(\chi-\chi_{0})}[1+e\cos(\psi-\psi_{0})]^{2}}\\
        &\cdot[e(12+4e^{2}-10p+p^{2})-2(6+2e^{2}-p)(-3+p)\\
        &\cdot\cos(\psi-\psi_{0})-e(6+2e^{2}-p)(-6+p)\cos^{2}(\psi-\psi_{0})\\
        &+2e^{2}(6+2e^{2}-p)\cos^{3}(\psi-\psi_{0})],
    \end{aligned}
\end{equation}
\begin{equation}
    \begin{aligned}
        C^{(e)}_{\phi}&=\frac{p\sqrt{(p-3-e^{2})(1-z_{1}^{2})}}{[(p-6)^{2}-4e^{2}][1+e\cos(\psi-\psi_{0})]^{2}}\\
        &\cdot [e(12+4e^{2}-10p+p^{2})-2(6+2e^{2}-p)(-3+p)\\
        &\cdot\cos(\psi-\psi_{0})-e(6+2e^{2}-p)(-6+p)\cos^{2}(\psi-\psi_{0})\\
        &+2e^{2}(6+2e^{2}-p)\cos^{3}(\psi-\psi_{0})],
    \end{aligned}
\end{equation}
\begin{equation}
    C^{(z_{1})}_{r}=0,
\end{equation}
\begin{equation}
    C^{(z_{1})}_{\theta}=\frac{p(1-z_{1}^{2})\sin(\chi-\chi_{0})\sqrt{p-3-e^{2}}}{\sqrt{1-z_{1}^{2}\cos^{2}(\chi-\chi_{0})}[1+e\cos(\psi-\psi_{0})]^{2}},
    \label{eq:Cz1theta}
\end{equation}
\begin{equation}
    C^{(z_{1})}_{\phi}=-\frac{pz_{1}\sin^{2}(\chi-\chi_{0})\sqrt{(p-3-e^{2})(1-z_{1}^{2})}}{[1+e\cos(\psi-\psi_{0})]^{2}}.
    \label{eq:Cz1phi}
\end{equation}
For $\psi_{0}$ and $\chi_{0}$, with
\begin{equation}
    \begin{aligned}
        \dv{\psi_{r}}{\tau}&=\dv{\psi}{\tau}-\dv{\psi_{0}}{\tau}=\dv{\psi_{r}}{\tau}\Big|_{\rm geodesic}-\dv{\psi_{0}}{\tau}\\
        &=\frac{1}{(\partial r/\partial\psi_{r})}\left(\dv{r}{\tau}-\pdv{r}{p}\dv{p}{\tau}-\pdv{r}{e}\dv{e}{\tau}\right),
    \end{aligned}
\end{equation}
and
\begin{equation}
    \begin{aligned}
        \dv{\psi_{\theta}}{\tau}&=\dv{\chi}{\tau}-\dv{\chi_{0}}{\tau}=\dv{\psi_{\theta}}{\tau}\Big|_{\rm geodesic}-\dv{\chi_{0}}{\tau}\\
        &=\frac{1}{(\partial\theta/\partial\psi_{\theta})}\left(\dv{\theta}{\tau}-\pdv{\theta}{z_{1}}\dv{z_{1}}{\tau}\right),
    \end{aligned}
\end{equation}
then
\begin{equation}
\begin{aligned}
    C^{(\psi_{0})}_{r}&=-\frac{(p-6)\cos(\psi-\psi_{0})+2e}{e[(p-6)^{2}-4e^{2}]}\\
    &\cdot\sqrt{p(p-3-e^{2})[p-6-2e\cos(\psi-\psi_{0})]},
\end{aligned}
\end{equation}
\begin{equation}
    \begin{aligned}
         C^{(\psi_{0})}_{\theta}&=-\frac{pz_{1}\sin(\chi-\chi_{0})\sin(\psi-\psi_{0})\sqrt{p-3-e^{2}}}{e[(p-6)^{2}-4e^{2}]\sqrt{1-z_{1}^{2}\cos^{2}(\chi-\chi_{0})}[1+e\cos(\psi-\psi_{0})]^{2}}\\
        &\cdot\left\{e[4e^{2}-(p-6)^{2}]\cos(\psi-\psi_{0})\right.\\
        &\left.+(-6+p)(6+e^{2}-2p+e^{2}\cos[2(\psi-\psi_{0})])\right\},
    \end{aligned}
\end{equation}
\begin{equation}
    \begin{aligned}
         C^{(\psi_{0})}_{\phi}&=-\frac{p\sqrt{(p-3-e^{2})(1-z_{1}^{2})}\sin(\psi-\psi_{0})}{e[(p-6)^{2}-4e^{2}][1+e\cos(\psi-\psi_{0})]^{2}}\\
        &\cdot\left\{e[4e^{2}-(p-6)^{2}]\cos(\psi-\psi_{0})\right.\\
        &\left.+(-6+p)(6+e^{2}-2p+e^{2}\cos[2(\psi-\psi_{0})])\right\},
    \end{aligned}
\end{equation}
\begin{equation}
    \begin{aligned}
        C^{(\chi_{0})}_{r}=0,
    \end{aligned}
\end{equation}
\begin{equation}
    \begin{aligned}
        C^{(\chi_{0})}_{\theta}=-\frac{p\sqrt{p-3-e^{2}}(1-z_{1}^{2})\cos(\chi-\chi_{0})}{z_{1}\sqrt{1-z_{1}^{2}\cos^{2}(\chi-\chi_{0})}[1+e\cos(\psi-\psi_{0})]^{2}},
    \end{aligned}
\end{equation}
\begin{equation}
    \begin{aligned}
        C^{(\chi_{0})}_{\phi}=\frac{p\sqrt{(p-3-e^{2})(1-z_{1}^{2})}\sin[2(\chi-\chi_{0})]}{2[1+e\cos(\psi-\psi_{0})]^{2}}.
    \end{aligned}
\end{equation}
\section{Adiabatic approximation form in this work}\label{appendix:average Integral2}
In this work, by explicitly accounting for $\psi_{0}$ and $\chi_{0}$, we adopt an equivalent form of Eq.~\eqref{eq:average integral1}, in which the phases $\psi$ and $\chi$ are integrated from $0$ to $2\pi$, as
\begin{equation}
    \expval{f}_{I_{0}}=\frac{\int_{0}^{2\pi}d\chi\int_{0}^{2\pi}d\psi\frac{1}{\omega_{r}(\psi_{r},\ \mathbf{I})}f(\psi_{r},\ \psi_{\theta},\ \mathbf{I})}{2\pi\int_{0}^{2\pi}d\psi\frac{1}{\omega_{r}(\psi_{r},\ \mathbf{I})}},
    \label{eq:average integral2}
\end{equation}
with $\psi_{r}=\psi-\psi_{0}$, $\psi_{\theta}=\chi-\chi_{0}$, $\mathbf{I}=\{p,\ e,\ z_{1}\}$ and $I_{0}=\{p,\ e,\ z_{1},\ \psi_{0},\ \chi_{0}\}$. 
\section{Effect of SMBH spin on orbital evolution}
\begin{figure}
    \centering
    \includegraphics[scale=0.6]{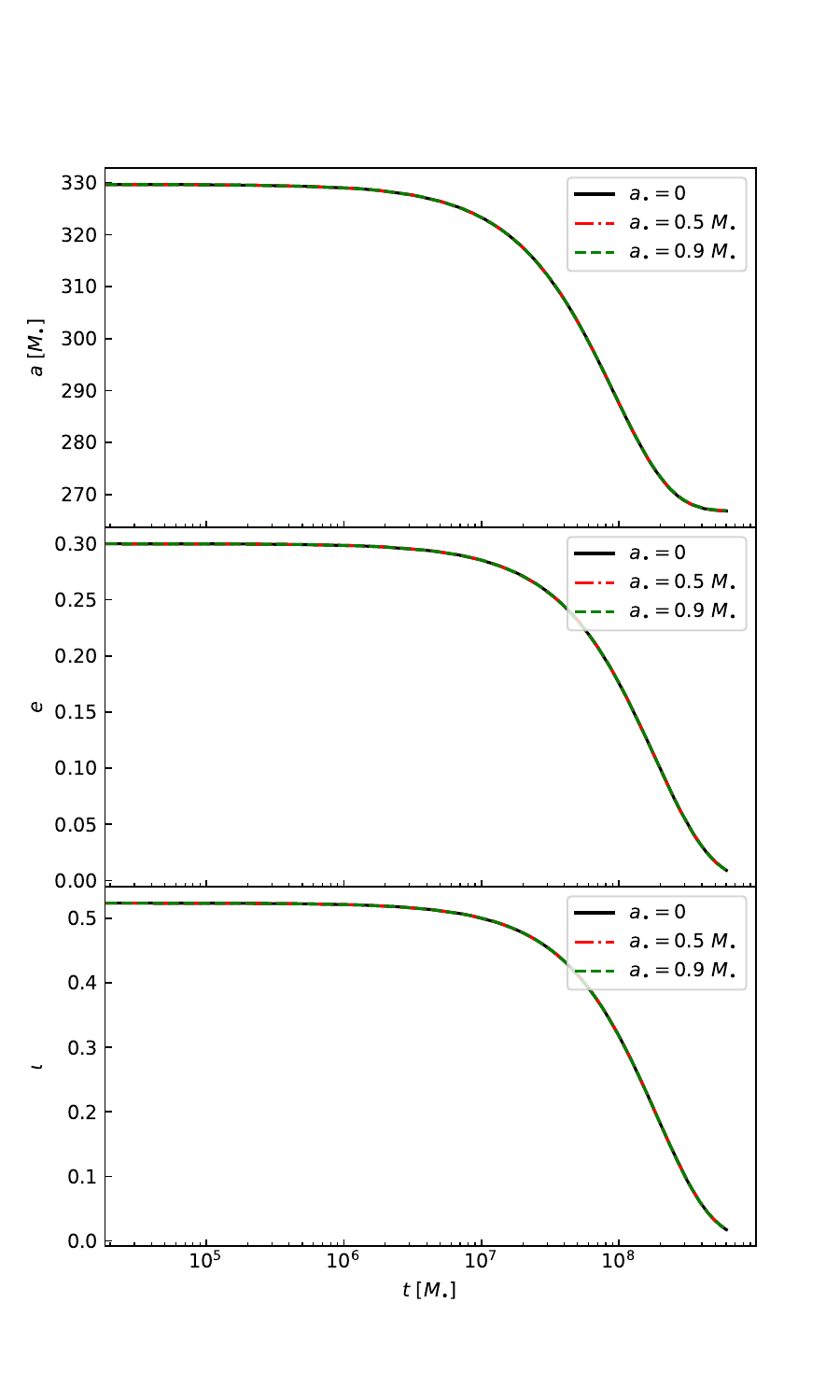}
    \caption{Comparison of the evolution of the semimajor axis $a$, eccentricity $e$, and orbital inclination angle $\iota$ for an EMRI system with central SMBH spin $a_{\bullet}=0$ (solid black lines), $0.5\ M_{\bullet}$ (dash-dot red lines)  and $0.9\ M_{\bullet}$ (dashed green lines), under the aero-drag force model. The initial parameters are $p_{\rm ini}=300\ M_{\bullet}$, $e_{\rm ini}=0.3$, $z_{1\ \rm ini}=0.5$, and $\gamma_{0}=2\times 10^{-6}\ M_{\bullet}^{-1}$.}
    \label{fig:spin contrast p_300_e_0_3_iota}
\end{figure}

In Fig.~\ref{fig:spin contrast p_300_e_0_3_iota}, we compare the orbital evolution of the SMO under the aero-drag force model for three representative cases with central SMBH spin $a_{\bullet}=0,\ 0.5\ M_{\bullet}$ and $0.9\ M_{\bullet}$. The trajectories of the key orbital parameters are found to be nearly indistinguishable among the three cases. This indicates that, within most of the parameter regimes considered in this work, the SMBH spin has a negligible impact on the capture process.
\section{Comparison with the gravitational-wave emission timescale}
\begin{figure}
    \centering
    \includegraphics[scale=0.6]{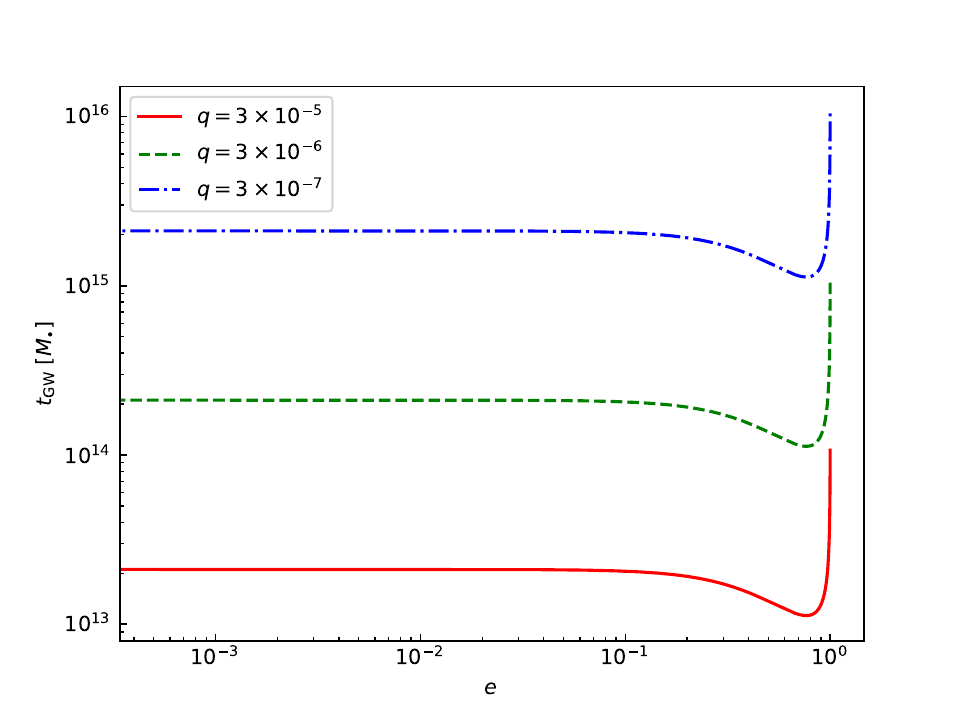}
    \caption{Estimation of the orbital shrinkage timescale $a/\abs{\expval{\mathrm{d}a/\mathrm{d}t}}$ as a function of the eccentricity $e$, for a fixed semilatus rectum $p=300\ M_{\bullet}$, and three representative mass ratios $q=3\times 10^{-5},\ 3\times 10^{-6}$, and $3\times 10^{-7}$.}
    \label{fig:t GW estiomation}
\end{figure}

In addition to interactions with disk, the orbital energy of EMRIs is also dissipated through gravitational-wave (GW) radiation~\cite{PhysRev.136.B1224,PhysRev.131.435}, The orbit-averaged decay rate of the semimajor axis is given by~\cite{PhysRev.136.B1224}
\begin{equation}
    \expval{\dv{a}{t}}=-\frac{64}{5}\frac{G^3 m_1 m_2(m_1 + m_2)}{c^5 a^3(1-e^2)^{\frac{7}{2}}}\left(1+\frac{73}{24}e^2+\frac{37}{96}e^4\right),
    \label{eq:GW a shrink average}
\end{equation}
where $m_1,\ m_2$ denote the primary and secondary masses, respectively, with $m_1\rightarrow M_{\bullet}$ and $m_2\rightarrow m_{\rm SMO}$ in this work. With the mass ratio $q\equiv m_{\rm SMO}/M_{\bullet}$, the corresponding GW emission timescale can be estimated as
\begin{equation}
    t_{\rm GW}=\frac{a}{\abs{\expval{\dv{a}{t}}}}=\frac{5}{64}\frac{a^4}{q(1+q)}\frac{(1-e^2)^{\frac{7}{2}}}{1+\frac{73}{24}e^2+\frac{37}{96}e^4}.
\end{equation}
The orbital shrinkage timescale due to GW emission is shown as a function of the eccentricity $e$ in Fig.~\ref{fig:t GW estiomation}. We adopt an SMO mass of $m_{\rm SMO}=30\ M_{\odot}$, consistent with Ref.~\cite{Wang2024}, and consider three representative SMBH masses, $M_{\bullet}=10^{6}\ M_{\odot},\ 10^{7}\ M_{\odot}$ and $10^{8}\ M_{\odot}$, corresponding to mass ratios $q=3\times 10^{-5},\ 3\times 10^{-6}$ and $3\times 10^{-7}$, respectively. Compared with the results obtained in this work, we find that $t_{\rm GW}$ is significantly longer than the evolution timescale associated with the aero-drag force model. For the dynamical friction model, $t_{\rm GW}$ can become comparable to the evolution timescale under certain initial conditions. However, we can still mainly focus on the SMO-disk interaction, since $t_{\rm GW}\propto a^4$, whereas $t_{\rm cap,\ dyn}\propto a^{-\frac{1}{2}}$ as summarized in Table~\ref{tab:power tau p}. For EMRIs with larger orbital separations, as considered in this work, this scaling implies $t_{\rm GW}\gg t_{\rm cap,\ dyn}$. Furthermore, to facilitate direct comparison with previous studies of disk-crossing processes performed within a Newtonian framework, we neglect the effects of GW radiation in our analysis.
\bibliographystyle{apsrev4-1}
\bibliography{reference}
\end{document}